\PassOptionsToPackage{unicode}{hyperref}
\PassOptionsToPackage{hyphens}{url}
\PassOptionsToPackage{dvipsnames,svgnames,x11names}{xcolor}
\documentclass[12pt]{article}
\usepackage{algorithm} 
\usepackage{algpseudocode}
\usepackage{amsmath,amssymb,amsthm}
\usepackage{bm}
\usepackage{tikz}
\newcommand*\circled[1]{\tikz[baseline=(char.base)]{
            \node[shape=circle,draw,inner sep=2pt] (char) {#1};}}
\usepackage{enumitem}
\usepackage{iftex}
\ifPDFTeX
  \usepackage[T1]{fontenc}
  \usepackage[utf8]{inputenc}
  \usepackage{textcomp} 
\else 
  \usepackage{unicode-math}
  \defaultfontfeatures{Scale=MatchLowercase}
  \defaultfontfeatures[\rmfamily]{Ligatures=TeX,Scale=1}
\fi
\usepackage{lmodern}
\ifPDFTeX\else  
\fi
\IfFileExists{upquote.sty}{\usepackage{upquote}}{}
\IfFileExists{microtype.sty}{
  \usepackage[]{microtype}
  \UseMicrotypeSet[protrusion]{basicmath} 
}{}
\makeatletter
\@ifundefined{KOMAClassName}{
  \IfFileExists{parskip.sty}{%
    \usepackage{parskip}
  }{
    \setlength{\parindent}{0pt}
    \setlength{\parskip}{6pt plus 2pt minus 1pt}}
}{
  \KOMAoptions{parskip=half}}
\makeatother
\usepackage{xcolor}
\makeatletter
\ifx\paragraph\undefined\else
  \let\oldparagraph\paragraph
  \renewcommand{\paragraph}{
    \@ifstar
      \xxxParagraphStar
      \xxxParagraphNoStar
  }
  \newcommand{\xxxParagraphStar}[1]{\oldparagraph*{#1}\mbox{}}
  \newcommand{\xxxParagraphNoStar}[1]{\oldparagraph{#1}\mbox{}}
\fi
\ifx\subparagraph\undefined\else
  \let\oldsubparagraph\subparagraph
  \renewcommand{\subparagraph}{
    \@ifstar
      \xxxSubParagraphStar
      \xxxSubParagraphNoStar
  }
  \newcommand{\xxxSubParagraphStar}[1]{\oldsubparagraph*{#1}\mbox{}}
  \newcommand{\xxxSubParagraphNoStar}[1]{\oldsubparagraph{#1}\mbox{}}
\fi
\makeatother

\usepackage{longtable,booktabs,array}
\usepackage{calc} 
\usepackage{etoolbox}
\makeatletter
\patchcmd\longtable{\par}{\if@noskipsec\mbox{}\fi\par}{}{}
\makeatother
\IfFileExists{footnotehyper.sty}{\usepackage{footnotehyper}}{\usepackage{footnote}}
\makesavenoteenv{longtable}
\usepackage{graphicx}
\makeatletter
\def\maxwidth{\ifdim\Gin@nat@width>\linewidth\linewidth\else\Gin@nat@width\fi}
\def\maxheight{\ifdim\Gin@nat@height>\textheight\textheight\else\Gin@nat@height\fi}
\makeatother
\setkeys{Gin}{width=\maxwidth,height=\maxheight,keepaspectratio}
\makeatletter
\def\fps@figure{htbp}
\makeatother

\makeatletter
\@ifpackageloaded{caption}{}{\usepackage{caption}}
\AtBeginDocument{%
\ifdefined\contentsname
  \renewcommand*\contentsname{Table of contents}
\else
  \newcommand\contentsname{Table of contents}
\fi
\ifdefined\listfigurename
  \renewcommand*\listfigurename{List of Figures}
\else
  \newcommand\listfigurename{List of Figures}
\fi
\ifdefined\listtablename
  \renewcommand*\listtablename{List of Tables}
\else
  \newcommand\listtablename{List of Tables}
\fi
\ifdefined\figurename
  \renewcommand*\figurename{Figure}
\else
  \newcommand\figurename{Figure}
\fi
\ifdefined\tablename
  \renewcommand*\tablename{Table}
\else
  \newcommand\tablename{Table}
\fi
}
\@ifpackageloaded{float}{}{\usepackage{float}}
\floatstyle{ruled}
\@ifundefined{c@chapter}{\newfloat{codelisting}{h}{lop}}{\newfloat{codelisting}{h}{lop}[chapter]}
\floatname{codelisting}{Listing}

\makeatother
\makeatletter
\@ifpackageloaded{caption}{}{\usepackage{caption}}
\@ifpackageloaded{subcaption}{}{\usepackage{subcaption}}
\makeatother

\ifLuaTeX
  \usepackage{selnolig}  
\fi
\usepackage[]{natbib}
\usepackage{bookmark}

\IfFileExists{xurl.sty}{\usepackage{xurl}}{} 
\hypersetup{
  pdftitle={Title},
  pdfauthor={Author 1; Author 2},
  pdfkeywords={3 to 6 keywords, that do not appear in the title},
  colorlinks=true,
  linkcolor={blue},
  filecolor={Maroon},
  citecolor={Blue},
  urlcolor={Blue},
  pdfcreator={LaTeX via pandoc}}

\newtheorem{theorem}{Theorem}
\newcommand{\dd}{\mbox{d}}
\newtheorem{assumption}{Assumption}

\newcommand{\anon}{1}

\begin{document}

\def\spacingset#1{\renewcommand{\baselinestretch}%
{#1}\small\normalsize} \spacingset{1}


\if1\anon
{
  \title{\LARGE\bf Robust Joint Modeling for Data with Continuous and Binary Responses}
\author{Yu Wang $^a$, Ran Jin $^b$, Lulu Kang $^a$\\
  \small
    $^a$ Department of Mathematics and Statistics, University of Massachusetts Amherst, \\
    \small
    Amherst, MA, United States \\
  \small
    $^b$ Grado Department of Industrial and Systems Engineering, Virginia Tech, \\
    \small
    Blacksburg, VA, United States 
     }
     \date{}

  \maketitle
} \fi
\if0\anon
{
  \bigskip
  \bigskip
  \bigskip
  \begin{center}
    {\LARGE\bf Robust Joint Modeling for Data with Continuous and Binary Responses}
\end{center}
  \medskip
} \fi
 
\bigskip
\begin{abstract}
In many supervised learning applications, the response consists of both continuous and binary outcomes. 
Studies have shown that jointly modeling such mixed-type responses can substantially improve predictive performance compared to separate analyses. 
But outliers pose a new challenge to the existing likelihood-based modeling approaches.  
In this paper, we propose a new robust joint modeling framework for data with both continuous and binary responses. 
It is based on the density power divergence (DPD) loss function with the $\ell_1$ regularization. 
The proposed framework leads to a sparse estimator that simultaneously predicts continuous and binary responses in high-dimensional input settings while down-weighting influential outliers and mislabeled samples. 
We also develop an efficient proximal gradient algorithm with Barzilai-Borwein spectral step size and a robust information criterion (RIC) for data-driven selection of the penalty parameters. 
Extensive simulation studies under a variety of contamination schemes demonstrate that the proposed method achieves lower prediction error and more accurate parameter estimation than several competing approaches. 
A real case study on wafer lapping in semiconductor manufacturing further illustrates the practical gains in predictive accuracy, robustness, and interpretability of the proposed framework.
\end{abstract}

\noindent%
\vfill

\newpage
\spacingset{1.8} 

\normalsize
\section{Introduction}\label{sec-intro}

In semiconductor manufacturing, the lapping process determines wafer flatness and thickness uniformity. 
This process generates two interdependent quality outcomes: a continuous response, such as total thickness variation (TTV), and a binary indicator, such as the site total indicator reading (STIR). Modeling these outcomes jointly is essential for improving manufacturing reliability. 
However, real-world data is often contaminated by measurement errors, sensor malfunctions, or mislabeled samples, which can lead to unstable model fitting in traditional frameworks.

\begin{figure}[htb]
\centering
\includegraphics[width=\linewidth]{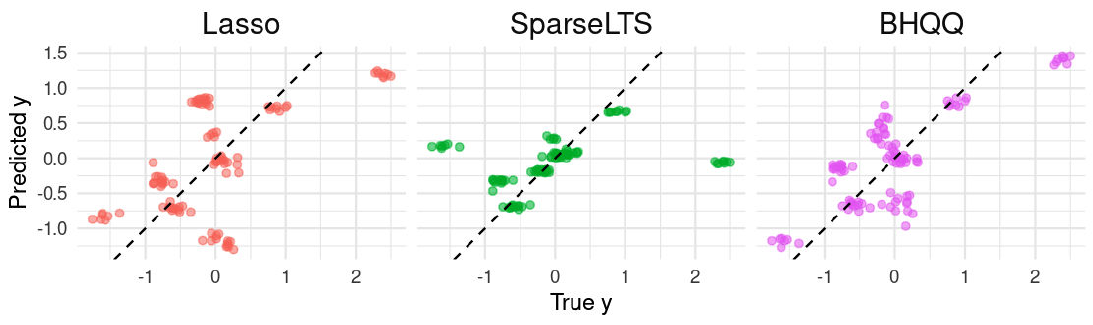}
\caption{Predicted TTV v.s. observations using three methods for the wafer lapping data.}
\label{fig:motivation}
\end{figure}

Figure \ref{fig:motivation} illustrates this sensitivity using a wafer lapping dataset \citep{Kang03072018}. 
Three representative methods—Lasso \citep{Tibshiranilasso}, SparseLTS \citep{Alfons2013SparseLTS}, and the Bayesian hierarchical quantitative and qualitative (BHQQ) model \citep{Kang03072018}—fail to accurately fit the data in the presence of outliers. 
While Lasso is sensitive to response contamination, SparseLTS lacks a joint modeling framework, and BHQQ lacks scalability and robustness against heavy-tailed noise. 
The visible deviations from the 45-degree reference line in Figure \ref{fig:motivation} highlight the need for a unified, robust joint modeling framework.

\subsection{Review of Existing Joint Modeling Methods}

In addition to the Bayesian hierarchical quantitative and qualitative model (BHQQ) of \cite{Kang03072018}, several studies have attempted to model continuous and binary responses jointly, each with its own limitations. 
\cite{Fitzmaurice2012} introduced a conditional regression model in which the continuous response is modeled conditional on the binary response, employing a logistic model for the qualitative component, but noted that the maximum likelihood estimation becomes considerably more complicated when the basic sampling unit is a cluster. 
Building on \cite{Fitzmaurice2012}, \cite{Lin2010Association} extended the joint modeling of clustered binary and continuous outcomes through a bivariate random-effects framework that allows both marginal and conditional interpretations, but it still lacks robustness to model misspecification and outliers. 
\cite{Craiu2012} developed a Bayesian conditional copula framework for modeling the dependence between binary and continuous outcomes using cubic splines and adaptive MCMC, offering flexible inference and copula selection through the DIC criterion. 
In manufacturing applications, such as the wafer lapping process, \cite{Deng15} proposed to jointly analyze quantitative and qualitative responses by capturing their inherent dependence through a combined likelihood framework and incorporating the nonnegative garrote penalty to achieve variable selection and improved prediction. 
Building on this line of work, \cite{Kang2022GAQQ} further developed a generative approach that jointly models quantitative and qualitative responses in high-dimensional settings by incorporating sparsity through penalized likelihood and establishing theoretical guarantees for consistency and asymptotic optimality.

While these existing models provide important insights into the relationship between quantitative and qualitative responses, they still have several limitations when applied to real-world data. 
In practice, the measurements collected from sensors or inspection systems often contain noise, mislabeled samples, or extreme values caused by equipment failure and process variation. 
Such contamination can lead to large estimation bias and unstable prediction, since most likelihood-based methods are highly sensitive to departures from model assumptions. 
Moreover, many existing joint modeling approaches are designed for moderate-dimensional data and may not scale well when the number of predictors is large. 
These challenges motivate the need for a robust and scalable modeling framework that can jointly handle continuous and binary responses under potential contamination.

Robust modeling has been an important topic in statistics and machine learning with a rich literature. 
Many estimators were designed to be resistant to deviations from idealized model assumptions. 
Seminal approaches include M-estimators (e.g., Huber loss by \cite{huber1964robust}), which bound the influence of outliers; R-estimators \citep{Heiler01011988}, based on ranks; and S-estimators \citep{10.1007/978-1-4615-7821-5_15}, which minimize a robust measure of scale of the residuals. 
\cite{https://doi.org/10.1002/wics.1524} gave a comprehensive review of robust regression. 
These estimators have been extended to generalized linear models combined with variable selection or regularization methods, including \cite{Khan01122007}, \cite{ALFONS2016421}, \cite{chang2018robust}, \cite{loh2017statistical}, etc. 

Recently, a new group of robust estimators has emerged from information theory: the use of divergences. 
Divergences, such as the Kullback-Leibler (KL) Divergence, measure the ``distance'' or discrepancy between two probability distributions. 
The fundamental connection to regression was established by recognizing that minimizing the OLS objective function is equivalent to minimizing the KL divergence between the assumed parametric model of the data and the true, underlying data-generating distribution. 
This insight opens the door to a generalized approach: replacing the KL divergence with a broader class of divergences that are inherently more robust to model misspecification.
Prominent examples of such robust divergences include: \cite{ghosh2013robust} introduced a Density Power Divergence (DPD) \citep{Basu1998robust} minimization for linear regression; \cite{chi2014robust} developed a robust penalized logistic regression algorithm based on a minimum distance criterion, which is a special case of DPD; \cite{alquier2023universal} built two estimators based on minimizing the Maximum Mean Discrepancy (MMD) \citep{sriperumbudur2011universality,sriperumbudur2010hilbert} that are both proven to be robust to Huber-type contamination. 

\subsection{Our Contribution}
Although the existing robust supervised learning approaches can overcome the effects of outliers to different degrees, some of them can also handle high-dimensional input; none of them can simultaneously model mixed types of responses. 
On the other hand, existing joint modeling approaches have not addressed outliers. 
To overcome these issues, we develop a new joint model for the continuous and binary responses based on the DPD from \cite{Basu1998robust}, as it is effective in regression \citep{ghosh2013robust} and classification \citep{chi2014robust} separately. 
The proposed framework applies the DPD loss for both types of outcomes instead of the traditional likelihood, which naturally down-weights the influence of outliers in both input variables and responses.
At the same time, an $\ell_1$ regularization is incorporated to achieve sparsity and enhance interpretability when dealing with high-dimensional input variables. 
We show that the proposed estimator under the DPD framework is consistent and asymptotically normal under mild regularity conditions, which enables statistical inference. 
Computationally, a proximal gradient algorithm with the Barzilai-Borwein step size is developed to efficiently solve the minimization of the DPD with $\ell_1$ penalty. 
The tuning parameters are selected through a robust information criterion (RIC), which balances model fit and complexity without being overly influenced by outliers. 
The results of the simulation examples and a real case study show that the proposed robust model consistently achieves better estimation and prediction accuracy than some existing approaches when input and output data are both contaminated. 
Overall, this study contributes a unified, theoretically justified, and computationally efficient framework for robust modeling of data with both continuous and binary responses. 

The remainder of the paper is organized as follows. 
In Section \ref{sec-meth}, we introduce the DPD loss function for the continuous and binary responses and the asymptotic property of the estimators. 
Section \ref{sec-optimization} describes the proximal-gradient algorithm and our procedure for selecting tuning parameters. 
Section \ref{sec-experiment} and \ref{sec:real} present simulation studies and a real case study of the lapping process. 
Section \ref{sec-conc} concludes the paper.

\section{DPD Loss}\label{sec-meth}

We consider the observed data $\{\bm x_i,\, y_i,\, z_i\}$ for $i=1,\ldots, n$, where $\bm x_i \in \mathbb{R}^p$ are the $p$-dimensional predictors, $y_i \in \mathbb{R}$ is the continuous response, and $z_i \in \{0,1\}$ is the binary response. 
To jointly model the continuous and binary responses $y$ and $z$ given $\bm x$, we express their joint probability density function as: $f(y,\,z \mid \bm x)=f(y\mid z,\,\boldsymbol{x})f(z \mid \bm x)$, where $f(\cdot)$ represents a general probability density function. 
As similarly in \cite{Deng15} and \cite{Kang03072018}, $z$ is modeled using logistic regression, while the conditional distribution of $y$ given $z$ and $\bm x$ follows a linear regression model.
Specifically, 
\begin{align*}
y \mid z, \boldsymbol{x} &\sim N\big(z \boldsymbol{x}^\top \boldsymbol{\beta} + (1-z)\boldsymbol{x}^\top \boldsymbol{\omega},\,\sigma^2\big), 
\end{align*}
where $\boldsymbol{\beta} = (\beta_1,\, \ldots,\, \beta_p )^\top$ and $\boldsymbol{\omega} = (\omega_1,\, \ldots,\, \omega_p )^\top$ are regression coefficients of $y$ for $z=1$ and $z=0$, respectively. 
We assume that $y$ conditional on $(z=1,\bm x)$ and $(z=0,\bm x)$ has the same variance $\sigma^2$ for simplification.
The marginal model of $z\mid \bm x$ is 
\begin{align*}
z \mid \boldsymbol{x} &=
\begin{cases}
1, & \text{w.p. } p(\boldsymbol{x}), \\
0, & \text{w.p. } 1 - p(\boldsymbol{x})
\end{cases} 
\quad \text{with } p(\boldsymbol{x}) = \frac{\exp(\boldsymbol{x}^\top \boldsymbol{\eta})}{1 + \exp(\boldsymbol{x}^\top \boldsymbol{\eta})}, 
\end{align*} 
where $\boldsymbol{\eta} = (\eta_1, \ldots, \eta_p)^\top$ are the logistic regression coefficients.  
In \cite{Deng15} and \cite{Kang03072018}, the joint likelihood function based on $f(y\mid z,\ boldsymbol{x})f(z \mid \bm x)$ is used for the next stage of the frequentist or Bayesian approach. 

The original density power divergence (DPD) proposed by \cite{Basu1998robust} is defined by 
\[
d_{\alpha}(g,f)=\int \left[f^{1+\alpha}(x)-\left(1+\frac{1}{\alpha}\right)f^{\alpha}(x)+\frac{1}{\alpha}g^{1+\alpha}(x)\right]\dd x, \text{ for }\alpha>0.
\]
Here $g$ and $f$ are two density functions. 
\cite{Basu1998robust} showed that $d_{\alpha}(g,f)$ is nonnegative and for all $g$ and $f$ in $\mathcal{D}$, where $\mathcal{D}$ is the class of all distributions having densities with respect to Lebesgue measure. 
The divergence is equal to zero if and only if $f\equiv g$ almost everywhere. 
When $\alpha=0$, the above definition $d_{\alpha}(g,f)$ is not defined. 
Instead, \cite{Basu1998robust} defined $d_0(g,f)$ as
\[d_0(g,f)=\lim_{\alpha\to 0} d_{\alpha}(g,f)=\int g(z)\log (g(x)/f(x))\dd x,\]
which is a version of the Kullback-Leibler divergence. 
When $\alpha=1$, it corresponds to the $L_2E$ measure \citep{scott2001parametric, chi2014robust}.
According to \cite{Basu1998robust}, a larger $\alpha$ yields greater robustness to outliers, while a smaller $\alpha$ leads to higher statistical efficiency. 

In our context, we focus on two random variables $(y,z)$ given the predictor vector $\bm x$, and the $d_{\alpha}(g,f)$ becomes 
\begin{equation}\label{eq:dpd}
d_\alpha(g,f)= \iint [f^{1+\alpha}(y,z)-(1+\frac{1}{\alpha})g(y,z) f^\alpha(y,z)+\frac{1}{\alpha} g^{1+\alpha}(y,z)] \dd(y,\,z), \text{ for } \alpha>0.
\end{equation}
The density functions $f$ and $g$ have the following interpretation: $f(y, z\mid \bm x)$ denotes the model-based density function, parameterized by $\bm \theta=(\bm \beta, \bm \omega, \bm \eta)$ and $\sigma^2$, representing the probability of observing data $(y, z)$ given covariates $\bm x$ under the assumed statistical model. 
Specifically,
\[f(y, z \mid \bm{x}) = \phi(y \mid z \bm{x}^\top \bm \beta + (1-z)\bm{x}^\top \bm \omega, \sigma^2) \, p(\bm{x} \mid \bm \eta)^z (1-p(\bm{x} \mid \bm\eta))^{1-z},\]
where $\phi(\cdot \mid \mu, \sigma^2)$ is the normal density with mean $\mu$ and variance $\sigma^2$ and $p(\bm{x} \mid \bm \eta) = \mathrm{logit}^{-1}(\bm{x}^\top \bm\eta)$ gives the Bernoulli probability for $z$.
Conversely, let $G$ represent the true, data-generating distribution, and let $g$ be the corresponding density function. 
Then, we substitute the $f$ and $g$ in \eqref{eq:dpd} with $f(y, z \mid \bm{x})$ and density $g(y, z \mid \bm{x})$, respectively. 

In practice, since $G$ and $g$ are unknown, $g$ has to be approximated by the empirical distribution for any given $\bm x$, which is still denoted by $g$ to simplify notation.
The DPD loss function applied to the training data is 
\begin{align*}
&Q_{\alpha}(\bm \theta, \sigma^2) = \frac{1}{n}\sum_{i=1}^n d_{\alpha} (g(\cdot\mid \bm x_i), f(\cdot\mid \bm x_i))\\
=&\frac{1}{n}\sum_{i=1}^n\iint f^{1+\alpha}(y,z\mid \bm x_i)\dd (y,z)-\frac{1}{n}\left(1+\frac{1}{\alpha}\right)\sum_{i=1}^n \iint g(y, z \mid \bm x_i)f^{\alpha}(y,z\mid \bm x_i)\dd y\dd z\\
&+\frac{1}{\alpha} \frac{1}{n}\sum_{i=1}^n g^{1+\alpha}(y_i, z_i\mid \bm x_i)=\circled{1}+\circled{2}+\circled{3}. 
\end{align*}
The third term $\circled{3}$ is a constant term with respect to the parameters and thus is dropped from the loss function.
Upon substituting the explicit expressions for $f$ and the empirical approximation for $g$ into the general DPD formulation, the first two terms in $Q_{\alpha}$ are
\begin{align*}
\circled{1} &=\frac{1}{n}\sum_{i=1}^n\iint f^{1+\alpha}(y,z \mid \bm{x}_i) \, \dd y \, \dd z\\
= & \frac{1}{n} \frac{1}{(2\pi\sigma^2)^{\alpha/2}} \frac{1}{\sqrt{1+\alpha}}\sum_{i=1}^n
   \Big[ p(\bm{x}_i\mid\bm{\eta})^{1+\alpha} +
        \big(1 - p(\bm{x}_i\mid\bm{\eta})\big)^{1+\alpha} \Big],
\end{align*}
\begin{align*}
\circled{2}&=\frac{1}{n}\left(1+\frac{1}{\alpha}\right)\sum_{i=1}^n\iint g(y , z \mid \bm x) f^\alpha(y , z \mid \bm x) \, \dd y \, \dd z \\
&= \frac{1}{n}\left(1+\frac{1}{\alpha}\right)\left(\frac{1}{\sqrt{2\pi\sigma^2}}\right)^{\!\alpha}\left(
   \sum_{i : z_i = 1}^{n_{z=1}}
   \exp\!\left( -\frac{\alpha (y_i - \bm{x}_i^\top \bm{\beta})^2}{2\sigma^2} \right) p^{\alpha}(\bm x_i\mid \bm \eta)\right.
\\
&\left.\quad + 
   \sum_{i : z_i = 0}^{n_{z=0}}
   \exp\!\left( -\frac{\alpha (y_i - \bm{x}_i^\top \bm{\omega})^2}{2\sigma^2} \right) 
   \left( 1-p(\bm x_i\mid \bm \eta) \right)^{\alpha}
   \right).
\end{align*}
Combining the first and second terms and omitting the irrelevant constant, we obtain the loss function for parameter estimation. 
The final DPD-based loss function is thus given by: 
\begin{align*}
& Q_\alpha(\bm \theta, \sigma^2) 
= \frac{1}{\sqrt{\alpha+1}} \frac{1}{n} 
\sum_{i=1}^n \Big[ p(\boldsymbol{x}_i \mid \boldsymbol{\eta})^{1+\alpha} 
+ \big(1 - p(\boldsymbol{x}_i \mid \boldsymbol{\eta})\big)^{1+\alpha} \Big] \\
& \quad 
- \left(1 + \frac{1}{\alpha}\right) \frac{1}{n}  
\sum_{i:\{z_i=1\}}^{n_{z=1}} 
\exp\!\left(-\frac{\alpha (y_i - \boldsymbol{x}_i^\top \boldsymbol{\beta})^2}{2\sigma^2}\right)
\big(p(\boldsymbol{x}_i \mid \boldsymbol{\eta})\big)^{\alpha}\\
& \quad 
- \left(1 + \frac{1}{\alpha}\right) \frac{1}{n}  
\sum_{i:\{z_i=0\}}^{n_{z=0}} 
\exp\!\left(-\frac{\alpha (y_i - \boldsymbol{x}_i^\top \boldsymbol{\omega})^2}{2\sigma^2}\right)
\big(1 - p(\boldsymbol{x}_i \mid \boldsymbol{\eta})\big)^{\alpha}.
\end{align*}
Details regarding the derivation of these terms can be found in the Supplement \ref{sec:Derivation_loss}. 

The DPD robust estimator $\hat{\bm \theta}$ is the minimizer of $Q_{\alpha}(\bm \theta, \sigma^2)$ for given $\sigma^2$. 
To investigate its theoretical properties, we show that it satisfies the five assumptions of Theorem 2.2 of \cite{Basu1998robust}  , which established that the general DPD robust estimator is consistent and has a multivariate normal asymptotic distribution. 
To keep the paper concise, we write the assumptions for the joint model and verify that they are satisfied. 
Consequently, the proposed DPD robust estimator also has the same consistent and asymptotic properties, summarized in Theorem \ref{thm:asympt}.

\begin{theorem}\label{thm:asympt}
Let $\hat{\bm \theta}_n=(\hat{\bm \beta}_n, \hat{\bm \omega}_n, \hat{\bm \eta}_n)$ be the minimizer of $Q_{\alpha}(\bm \theta, \sigma^2)$ for given $\sigma^2$. 
Under the Assumptions A.1-A.5 stated in \ref{sec:assump},  with probability equal to 1 and as $n\to \infty$, there exists $(\hat{\bm \beta}_n, \hat{\bm \omega}_n, \hat{\bm \eta}_n)$ such that 
\begin{enumerate}
\item $(\hat{\bm \beta}_n, \hat{\bm \omega}_n, \hat{\bm \eta}_n)$ is consistent for $(\bm \beta, \bm \omega, \bm \eta)$, 
\item $(\hat{\bm \beta}_n, \hat{\bm \omega}_n, \hat{\bm \eta}_n)$ is asymptotically multivariate normal, i.e, 
\begin{align*}
n^{1/2}
\begin{bmatrix}
\hat{\boldsymbol{\beta}}_n - \boldsymbol{\beta} \\
\hat{\boldsymbol{\omega}}_n - \boldsymbol{\omega} \\
\hat{\boldsymbol{\eta}}_n - \boldsymbol{\eta}
\end{bmatrix}
&\stackrel{d}{\longrightarrow} \mathcal{MVN} \left(0, \bm J^{-1} \bm K \bm J^{-1} \right), \text{ as }n\to \infty. 
\end{align*}
\end{enumerate}
\end{theorem}
The explicit forms of each block of $\bm J$ and $\bm K$ involve lengthy algebraic expressions. 
For readability, we present the full derivations and closed-form expressions in Supplement \ref{sec:restate}. 

\section{Optimization Algorithm} \label{sec-optimization}
In this section, we introduce the proposed methods for parameter estimation under the DPD framework. To further improve robustness and facilitate variable selection, we incorporate $\ell_1$ regularization, leading to the final objective function:
\begin{align} \label{eq:minQ}
    h(\boldsymbol{\beta}, \boldsymbol{\omega}, \boldsymbol{\eta}; \sigma^2)
    = Q_\alpha(\boldsymbol{\beta}, \boldsymbol{\omega}, \boldsymbol{\eta}; \sigma^2)
    + \lambda_1 \lVert \boldsymbol{\beta} \rVert_1
    + \lambda_2 \lVert \boldsymbol{\omega} \rVert_1
    + \lambda_3 \lVert \boldsymbol{\eta} \rVert_1,
\end{align}
where $\lambda_1$, $\lambda_2$, and $\lambda_3$ are regularization parameters controlling the sparsity of the parameter estimates. We assume that $\boldsymbol{\beta}$, $\boldsymbol{\omega}$, and $\boldsymbol{\eta}$ are sparse, and estimate them by solving the $\ell_1$-regularized minimization problem in~\eqref{eq:minQ}.

\subsection{Estimation of the Variance}
In the proposed DPD joint modeling framework, the primary parameter of interest is $\bm \theta=(\bm \beta, \bm \omega, \bm \eta)$ and $\sigma^2$ is a nuisance parameter. 
Yet, the value of $\sigma^2$ is needed to evaluate the loss function and affects the accuracy of the $\bm \theta$ estimate. 
One could treat $\sigma^2$ as an unknown parameter and solve for it iteratively within the proximal gradient algorithm. 
However, our simulation study based on this idea showed that the estimated $\sigma^2$ lacked stability and was far from the true value when the data were heavily contaminated. 
Therefore, we adopt a plug-in strategy where a robust pilot estimate of $\sigma^2$ is determined before the main optimization. 
Specifically, we use the Pseudo Standard Error (PSE) as discussed in \cite{wu2009experiments}. 
The procedure for calculating the fixed $\sigma^2$ used in Algorithm \ref{alg:proximal} is as follows. 
We begin with a preliminary fit to the continuous response $y$ (independent of the binary response) using standard Lasso regression to obtain an initial set of residuals $r_{init}$. 
Although Lasso is sensitive to outliers, the subsequent PSE calculation is designed to filter out their influence. 
We first calculate an initial scale factor $s_0 = 1.5 \cdot \text{median}_i |r_{\text{init}, i}|$, and then identify a ``safe'' subset of residuals where $|r_{\text{init}, i}| < 2.5s_0$ to eliminate the impact of extreme outliers. 
The final robust scale estimate is defined as $\sigma = 1.5 \cdot \text{median}_{(|r_{\text{init}, i}|< 2.5s_0)} |r_{\text{init}, i}|$, where we enforce a lower bound $\epsilon = 10^{-12}$ for $\sigma^2$ to ensure numerical stability in experiment settings. 
This fixed $\sigma^2$ is then substituted into the objective function \eqref{eq:minQ} as a constant.
Once the optimization is finished, we use the estimated $(\hat{\bm \beta}_n, \hat{\bm \omega}_n)$ to obtain the final version of $\hat{\sigma}_2$. 

\subsection{Proximal Gradient Algorithm for Optimization Problem}\label{sec:PGA}
To solve the minimization problem, we apply the proximal gradient algorithm, which is particularly effective for $\ell_1$-regularized problems in high-dimensional settings. 
We follow the implementation of the proximal gradient descent for the sparse nonlinear regression under nonconvexity by \cite{yang2016sparse}, building on the foundational work of \cite{wright2009sparse}.
We combine the proximal gradient with the block coordinate descent and iteratively update sequence for the parameters $\boldsymbol{\beta}$, $\boldsymbol{\omega}$, and $\boldsymbol{\eta}$:
\begin{align}\label{eq:proximal1}
\boldsymbol{\beta}^{(t+1)} 
&= \arg \min_{\boldsymbol{\beta} \in \mathbb{R}^p} 
\Big\{ 
\langle \nabla_{\boldsymbol{\beta}} Q_\alpha( \boldsymbol{\beta}^{(t)}, \boldsymbol{\omega}^{(t)}, \boldsymbol{\eta}^{(t)}), 
\boldsymbol{\beta} - \boldsymbol{\beta}^{(t)} \rangle 
+ \frac{\gamma_{1,t}}{2} \lVert \boldsymbol{\beta} - \boldsymbol{\beta}^{(t)} \rVert_2^2 
+ \lambda_1 \lVert \boldsymbol{\beta} \rVert_1 
\Big\}\\
\label{eq:proximal2}
\boldsymbol{\omega}^{(t+1)} 
&= \arg \min_{\boldsymbol{\omega} \in \mathbb{R}^p} 
\Big\{ 
\langle \nabla_{\boldsymbol{\omega}} Q_\alpha( \boldsymbol{\beta}^{(t)}, \boldsymbol{\omega}^{(t)}, \boldsymbol{\eta}^{(t)}), 
\boldsymbol{\omega} - \boldsymbol{\omega}^{(t)} \rangle 
+ \frac{\gamma_{2,t}}{2} \lVert \boldsymbol{\omega} - \boldsymbol{\omega}^{(t)} \rVert_2^2 
+ \lambda_2 \lVert \boldsymbol{\omega} \rVert_1 
\Big\}\\
\label{eq:proximal3}
\boldsymbol{\eta}^{(t+1)} 
&= \arg \min_{\boldsymbol{\eta} \in \mathbb{R}^p} 
\Big\{ 
\langle \nabla_{\boldsymbol{\eta}} Q_\alpha( \boldsymbol{\beta}^{(t)}, \boldsymbol{\omega}^{(t)}, \boldsymbol{\eta}^{(t)}), 
\boldsymbol{\eta} - \boldsymbol{\eta}^{(t)} \rangle 
+ \frac{\gamma_{3,t}}{2} \lVert \boldsymbol{\eta} - \boldsymbol{\eta}^{(t)} \rVert_2^2 
+ \lambda_3 \lVert \boldsymbol{\eta} \rVert_1 
\Big\}
\end{align}
where $1/\gamma_{1,t}, 1/\gamma_{2,t}, 1/\gamma_{3,t} > 0$ are the step sizes at the $t$-th iteration. 
The gradient of $Q_{\alpha}$ is derived in Supplement \ref{sec:gradient}. 
By introducing the auxiliary variables 
$\boldsymbol{u}_1^{(t)} := \boldsymbol{\beta}^{(t)} - \frac{1}{\gamma_{1,t}} \nabla Q_{\beta}(\boldsymbol{\beta}^{(t)}, \boldsymbol{\omega}^{(t)}, \boldsymbol{\eta}^{(t)})$, 
$\boldsymbol{u}_2^{(t)} := \boldsymbol{\omega}^{(t)} - \frac{1}{\gamma_{2,t}} \nabla Q_{\omega}(\boldsymbol{\beta}^{(t)}, \boldsymbol{\omega}^{(t)}, \boldsymbol{\eta}^{(t)})$, 
and 
$\boldsymbol{u}_3^{(t)} := \boldsymbol{\eta}^{(t)} - \frac{1}{\gamma_{3,t}} \nabla Q_{\eta}(\boldsymbol{\beta}^{(t)}, \boldsymbol{\omega}^{(t)}, \boldsymbol{\eta}^{(t)})$, the update steps for $\boldsymbol{\beta}$, $\boldsymbol{\omega}$ and $\boldsymbol{\eta}$ can be reformulated from~\eqref{eq:proximal1},\eqref{eq:proximal2} and \eqref{eq:proximal3} as the following: 
\begin{align*}
\boldsymbol{\beta}^{(t+1)} & = \arg\min_{\boldsymbol{\beta} \in \mathbb{R}^p}
\left\{
\frac{1}{2}\lVert \boldsymbol{\beta} - \boldsymbol{u}_1^{(t)} \rVert_2^2
+ \frac{\lambda_1}{\gamma_{1,t}} \lVert \boldsymbol{\beta} \rVert_1
\right\}, \\
\boldsymbol{\omega}^{(t+1)} &= \arg\min_{\boldsymbol{\omega} \in \mathbb{R}^p}
\left\{\frac{1}{2}\lVert \boldsymbol{\omega} - \boldsymbol{u}_2^{(t)} \rVert_2^2 + \frac{\lambda_2}{\gamma_{2,t}} \lVert \boldsymbol{\omega} \rVert_1\right\},\\
\boldsymbol{\eta}^{(t+1)}  &= \arg\min_{\boldsymbol{\eta} \in \mathbb{R}^p}
\left\{
\frac{1}{2}\lVert \boldsymbol{\eta} - \boldsymbol{u}_3^{(t)} \rVert_2^2
+ \frac{\lambda_3}{\gamma_{3,t}} \lVert \boldsymbol{\eta} \rVert_1
\right\}.
\end{align*}
Each subproblem admits a closed-form solution via the soft-thresholding operator:
\begin{equation*}
\beta_i^{(t+1)} = \text{soft}\!\left(u_{1,i}^{(t)}, \frac{\lambda_1}{\gamma_{1,t}}\right), 
\omega_i^{(t+1)} = \text{soft}\!\left(u_{2,i}^{(t)}, \frac{\lambda_2}{\gamma_{2,t}}\right),
\eta_i^{(t+1)} = \text{soft}\!\left(u_{3,i}^{(t)}, \frac{\lambda_3}{\gamma_{3,t}}\right), \quad i = 1,\ldots,p .
\end{equation*}
where $\text{soft}:= \text{sign}(u)\max\{|u|-a,0\}$ denotes the soft-thresholding operator. Algorithm \ref{alg:proximal}  summarizes the overall procedure, which applies the SpaRSA technique proposed by \cite{wright2009sparse} to address the non-convexity of the optimization problem. The algorithm iteratively increases the step size $\gamma_{1,t},\gamma_{2,t},\gamma_{3,t}$ by a factor of $\zeta$ until the convergence criterion is satisfied, which guarantees a sufficient decrease of the objective function. 
To select the step sizes in each iteration, we utilize the Barzilai-Borwein spectral line search method from \cite{barzilai1988two}, as detailed in Algorithm \ref{alg:bb}.
Due to limited space, Algorithms \ref{alg:proximal} and \ref{alg:bb} are moved to Supplement \ref{sec:alg}.  
The settings of tuning parameters of both algorithms for all simulations and case studies are listed in Table \ref{tab:tuning_params} in Supplement \ref{sec:tuning}. 
 
\subsection{Framework for Selecting Penalty Parameter Based on RIC}

For data with outliers, conventional likelihood-based model selection criteria, such as AIC or BIC, are not ideal. 
In \cite{Basu1998robust}, a robust information criterion (RIC) was introduced for model selection. For each candidate model $M$, the RIC is defined as:
\begin{align}
    \text{RIC}_M 
    = Q_\alpha(\hat{\boldsymbol{\beta}}_{n,M}, \hat{\boldsymbol{\omega}}_{n,M}, \hat{\boldsymbol{\eta}}_{n,M})
    + (1+\alpha)\frac{\operatorname{tr}(\hat{\bm J}_M^{-1} \hat{\bm K}_M)}{n},
\end{align}
where $Q_\alpha(\hat{\boldsymbol{\beta}}_{n,M}, \hat{\boldsymbol{\omega}}_{n,M}, \hat{\boldsymbol{\eta}}_{n,M})$ denotes the DPD-based loss function evaluated at the estimated parameters, and the second term penalizes model complexity. 
To select the optimal regularization parameters $(\lambda_1, \lambda_2, \lambda_3)$, we employ the grid search based on the RIC in the 3-dim rectangular box whose lower and upper bounds $(\lambda_{i, \min}, \lambda_{i,\max})$ for $i=1,2, 3$ are set following the strategy by \cite{JSSv033i01}. 
Based on the grid search, we choose the best $(\lambda_1, \lambda_2, \lambda_3)$ returning the smallest RIC. 

\section{Numerical Experiments}\label{sec-experiment}

We consider two main simulation settings: small and large $p$ relative to the training sample size, where $p$ is the number of predictors. 
For the small-$p$ case, we let $p=8$ and the training sample size $n_{\text{train}}=700$, with an additional independent test set of size $n_{\text{test}}=300$. 
For the large-$p$ case, we set $p=50$, whereas the sample sizes of the training and testing sets remain the same as for $p=8$. 
In each simulation, new training and test data sets are generated, and for each combination of settings, we perform $B=100$ simulations. 
We evaluate the proposed DPD-estimator against three types of existing methods: regression-only, classification-only, and joint-modeling methods.
For regression-only methods, we consider: 
\begin{itemize}[leftmargin=*]
\item Lasso: the Lasso estimator of \cite{Tibshiranilasso}, implemented by the \texttt{glmnet} R package \citep{glmnet}; 
\item SparseLTS: the sparse least trimmed squares regression of \cite{Alfons2013SparseLTS}, implemented by the \texttt{robustHD} R package \citep{robustHD};
\item Lasso-QR: the Lasso-penalized quantile regression with $\tau=0.5$ by \cite{Koenker2005} and \cite{Belloni2011}, implemented by the \texttt{quantreg} R package \citep{quantreg}; 
\item Ada-LAD-Lasso: the adaptive least absolute deviation Lasso of \cite{Qin2017MTE}, implemented by the \texttt{MTE} R package \citep{MTE}.
 \end{itemize}
For the classification-only method, we again use the logistic regression from \texttt{glmnet}.
For joint-modeling both continuous and binary responses, we compare our method with the Bayesian hierarchical quantitative and qualitative model by \cite{Kang03072018}. 
The codes of BHQQ were obtained from the authors. 

We use three metrics to compare the performances. 
For the continuous response, we use the root mean squared prediction error on the test data set, $\text{RMSPE}=\frac{1}{m}\sum_{i=1}^m \sqrt{(y_i-\hat{y}_i)^2}$. 
For the binary response, we use the Misclassification Error on test data set, $\text{ME} = \frac{1}{m} \sum_{i=1}^nm\mathbf{1}(z_i \neq \hat{z}_i)$, where $\mathbf{1}(\cdot)$ is an indicator function. 
We also measure the parameter estimation accuracy by the $\ell_2$-norm errors for the parameters 
$\boldsymbol{\beta}$, $\boldsymbol{\omega}$, and $\boldsymbol{\eta}$: 
$\left\Vert \hat{\boldsymbol{\beta}} - \boldsymbol{\beta}^{\text{true}} \right\Vert$, 
$\left\Vert \hat{\boldsymbol{\omega}} - \boldsymbol{\omega}^{\text{true}} \right\Vert$, and
$\left\Vert \hat{\boldsymbol{\eta}} - \boldsymbol{\eta}^{\text{true}} \right\Vert$.

\subsection{Numerical Results for $p=8$}
Predictors $\boldsymbol{x}$ are independently generated from a multivariate normal distribution $\mathbf{\mathcal{MVN}}_p(\mathbf{0},\mathbf{I}_p)$. 
The response data are generated according to the model framework in Section \ref{sec-meth}, with a key deviation: the noise for the continuous response is generated from a Laplace distribution with mean 0 and standard deviation 1, rather than a Gaussian distribution. 
This deliberate violation of the model's distributional assumptions presents an additional challenge to statistical modeling. 
The true parameter vectors are $\bm{\beta}^\text{true}=(3,3,3,3,3,3,3,3)$, $\bm{\omega}^\text{true}=(5,5,5,5,5,5,5,5)$ and $\bm{\eta}^\text{true}=(5,5,5,5,5,5,5,5)$. 

We also introduce artificial contamination to the training data. 
For the predictors $\bm X$, the selected contaminated samples are replaced with samples from a multivariate normal distribution whose mean is shifted by +5 from the original sample means and whose covariance matrix is scaled by a factor of 1.2, simulating extreme outliers.
For $\bm Y$, the contaminated $y_i$'s are replaced by samples from $\mathcal{N}(20, 1)$. 
For $\bm Z$, the contaminated $z_i$'s are flipped (0 becomes 1 and vice versa). 
Using this alternative data generation schemes, we add the contamination in three ways: (1) one-way contamination: random select $15\%$ of the data to contaminate in $\bm X$, $\bm Y$, or $\bm Z$ alone; (2) two-way joint contamination: randomly select $15\%$ of data entries and contaminate both components in the pairs $(\bm X,\bm Y)$, $(\bm X, \bm {Z})$, or $(\bm Y, \bm Z  )$; (3) three-way joint contamination: randomly select $15\%$ of data entries and contaminate all three components $(\bm X,\bm Y,\bm Z)$ simultaneously.
These scenarios realistically emulate complex data corruption encountered in practice, presenting significant challenges for robust estimation and classification.

\begin{figure}[htbp]
    \centering
    \includegraphics[width=1\linewidth]{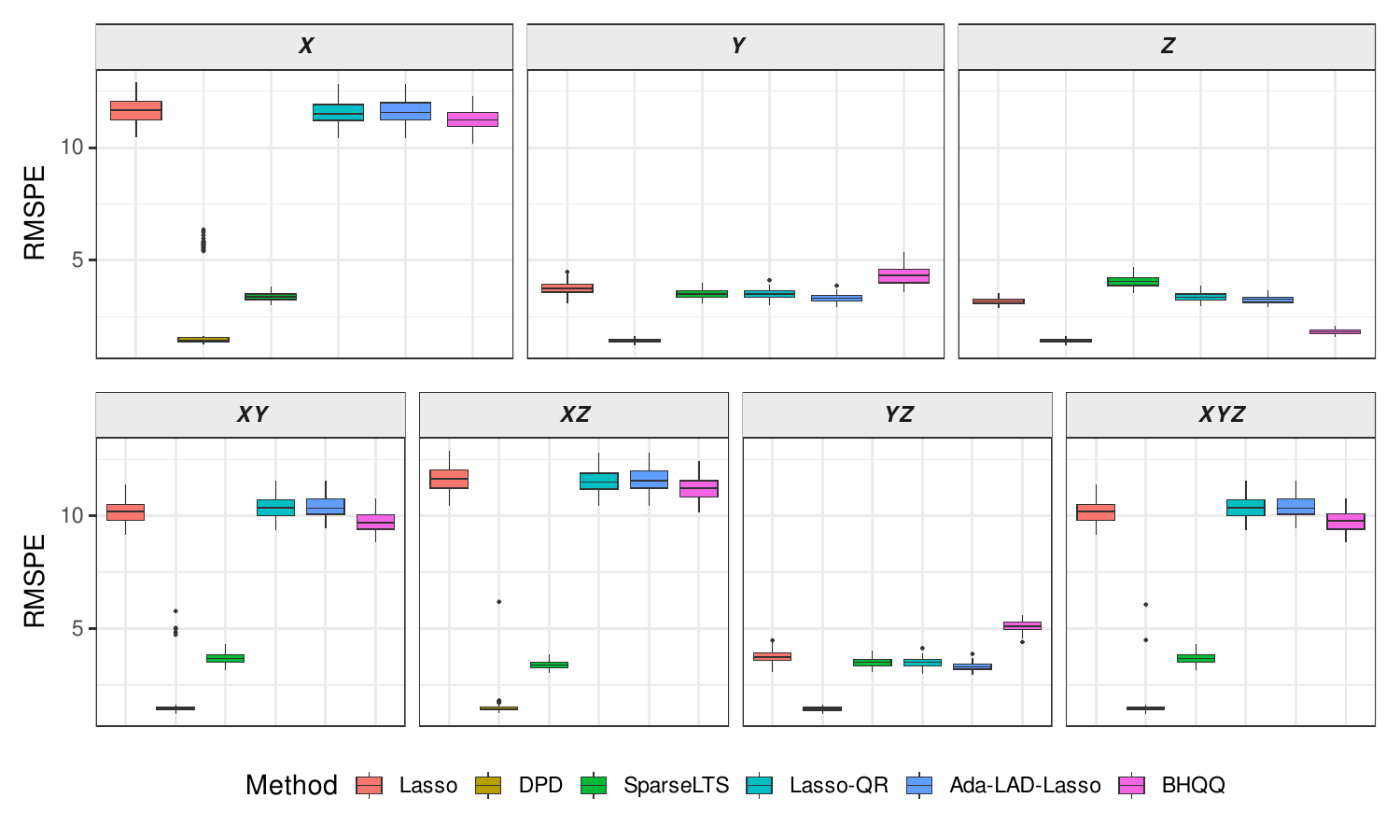}
    \caption{Boxplots of RMSPE from $B=100$ simulations with varying contamination type.}
    \label{fig:rmse_low}
\end{figure}

\begin{figure}[htbp]
    \centering
    \includegraphics[width=1\linewidth]{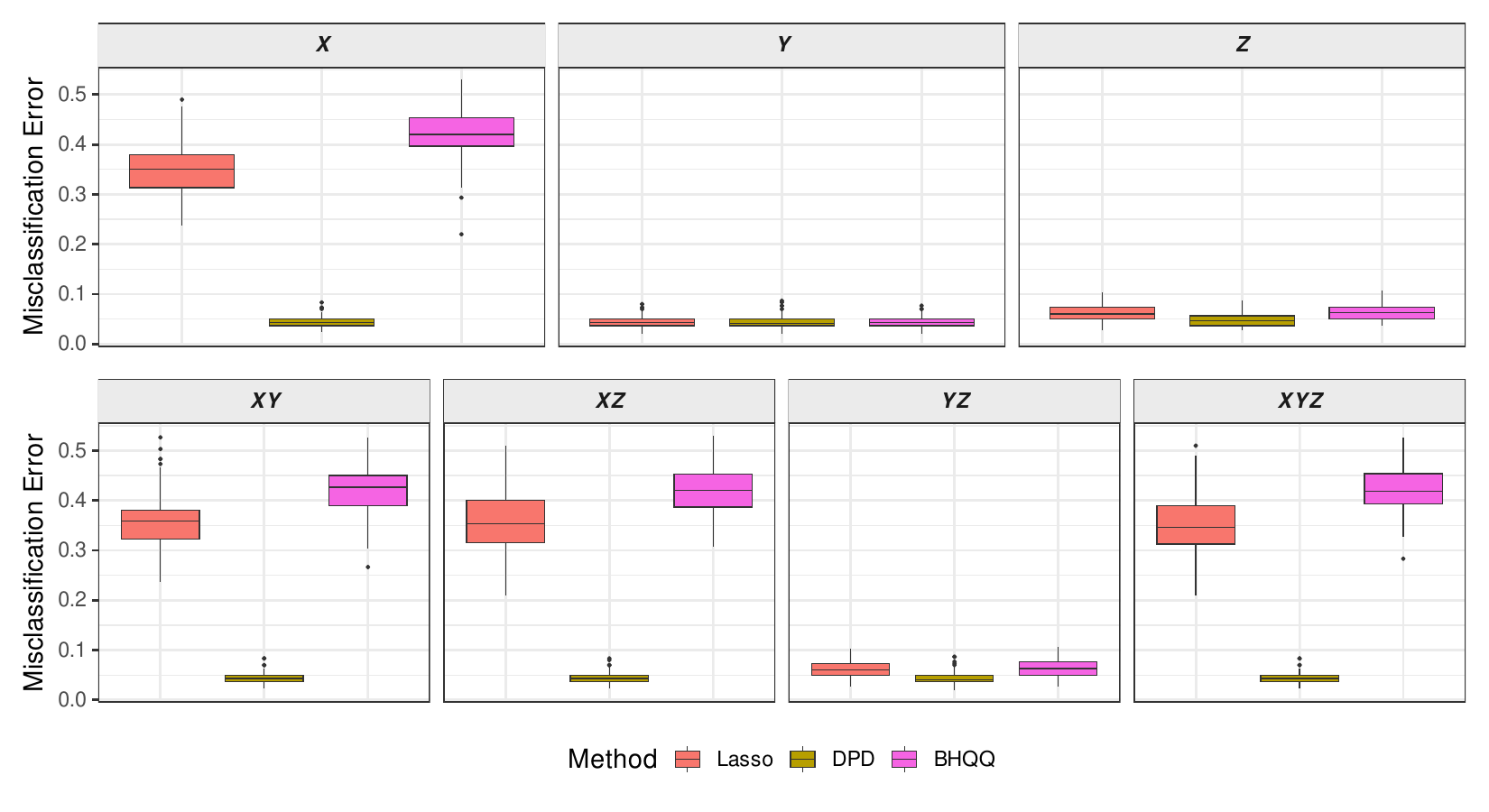}
    \caption{Boxplots of MEs from $B=100$ simulations with varying contamination type.}
    \label{fig:me_low}
\end{figure}

\begin{figure}[htbp]
    \centering
    \includegraphics[width=1.45\linewidth,  angle=90]{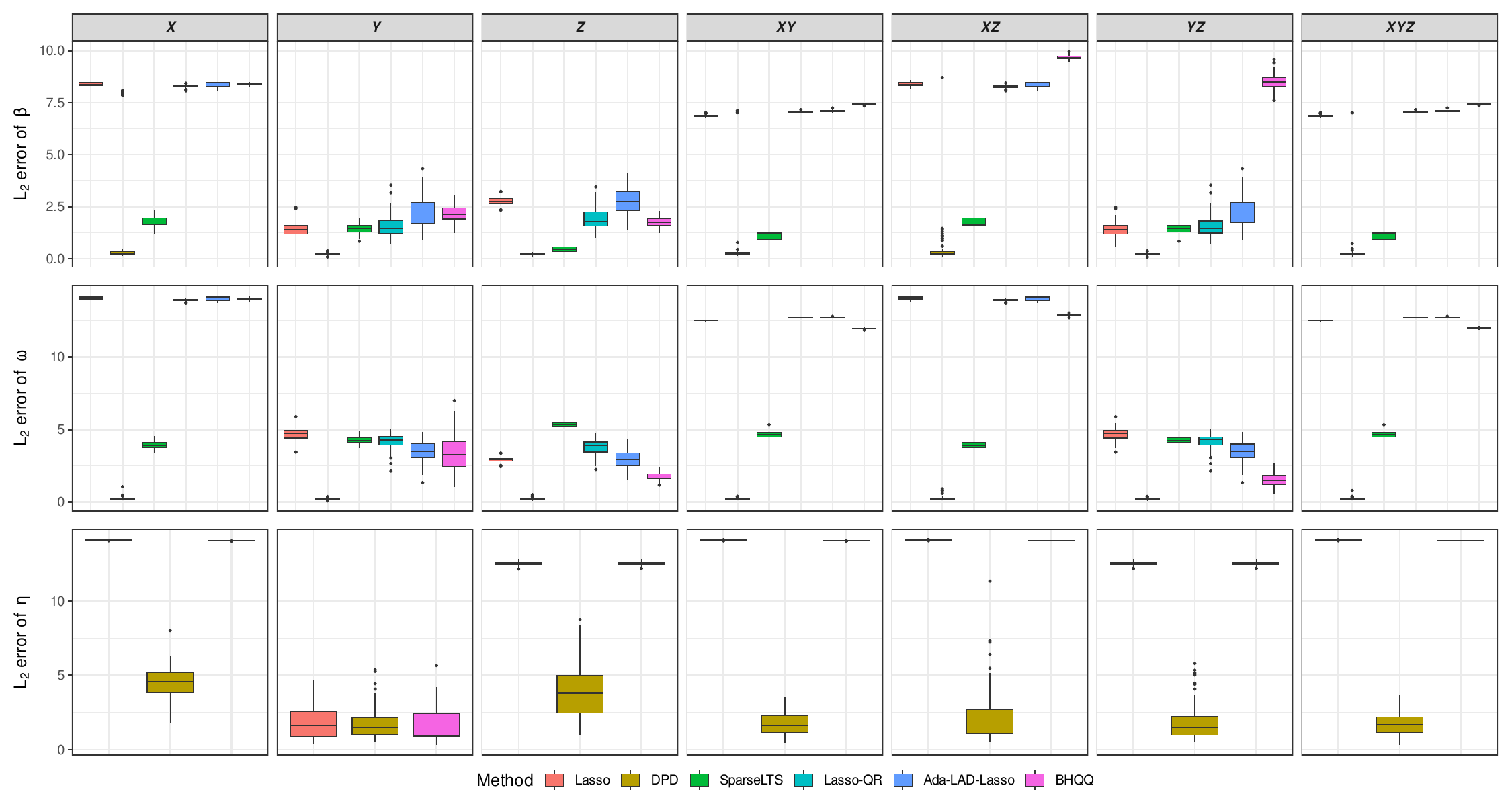}
    \caption{Boxplots of $\ell_2$ error for coefficient from $B=100$ simulations with varying contamination type.}
    \label{fig:l2e_low}
\end{figure}

The simulation results are summarized in Figure \ref{fig:rmse_low}, \ref{fig:me_low}, and s Supplemental \ref{fig:l2e_low}. 
The proposed DPD-based joint modeling approach consistently outperforms competing methods in parameter estimation accuracy, achieving the smallest $\ell_2$-norm errors for $\boldsymbol{\beta}$, $\boldsymbol{\omega}$, and $\boldsymbol{\eta}$ in nearly all contamination scenarios.
A notable exception occurs when only $\bm Y$ is contaminated, where our method yields a similar $\ell_2$-error for $\bm \eta$ compared to Lasso and BHQQ. 
This is expected for two reasons. 
First, the binary response $\bm Z $ itself contains no outliers in this scenario. 
Second, the DPD approach for the binary response is based solely on its marginal distribution, without leveraging information from the continuous response. 
Consequently, even in the absence of contamination in $\bm{Z}$, the DPD-based estimator maintains a clear advantage over other approaches, although the efficiency gap is somewhat narrowed.
In terms of ME, the proposed method is superior in all but the $\bm Y$-only contamination case, which aligns with the explanation above. 
For RMSPE, the DPD method generally outperforms others.

Overall, the DPD-based joint model demonstrates superior accuracy in parameter estimation across diverse contamination types. 
It is the only method capable of jointly modeling mixed-type outcomes while maintaining robustness to outliers in predictors and in either response type. 
In practice, we recommend performing multiple trials with different training-testing splits and selecting the model that yields the best prediction accuracy.
 

\spacingset{1.8}

\subsection{Numerical Results for $p = 50$}
For the case of $p=50$, $n_{\text{train}}=700$, and $n_{\text{test}}=300$, we set the simulation in a similar way as in \cite{Deng15}. 
The sparsity level $s\in\{0.1,0.2,0.5\}$ denotes the proportion of nonzero coefficients in each of $\bm \beta$, $\bm \omega$, and $\bm \eta$.
Let $S_t\subset\{1,\dots,p\}$ be the index set of the active predictors, and it is randomly selected with $|S_t|=\lfloor s\,p\rfloor$.
The three sets, $S_t(\bm \beta)$, $S_t(\bm \omega)$, and $S_t(\bm \eta)$, are drawn independently, so overlaps among the three active sets are possible and allowed. 
True coefficients are then generated as
\[
(\beta_t)_{t\in S_t(\bm \beta)}\overset{\text{i.i.d.}}{\sim}\mathcal N(3,1),\quad
(\omega_t)_{j\in S_t(\bm \omega)}\overset{\text{i.i.d.}}{\sim}\mathcal N(-5,1),\quad
(\eta_t)_{j\in S_t(\bm \eta)}\overset{\text{i.d.}}{\sim}\mathrm{Unif}(2,5),
\]
and all inactive parameters are set to zero.

Here, we follow the same data generation and contamination scheme as in the previous subsection. 
However, instead of fixing the level of contamination at $15\%$, we vary it at $5\%$, $10\%$, $15\%$, and $20\%$, enabling a systematic evaluation of method robustness. 
Also, we only use the three-way joint contamination, i.e., outliers are introduced jointly on the same subset of data entries across all inputs and outputs.

\begin{figure}[htbp]
    \centering
    \includegraphics[width=1\linewidth]{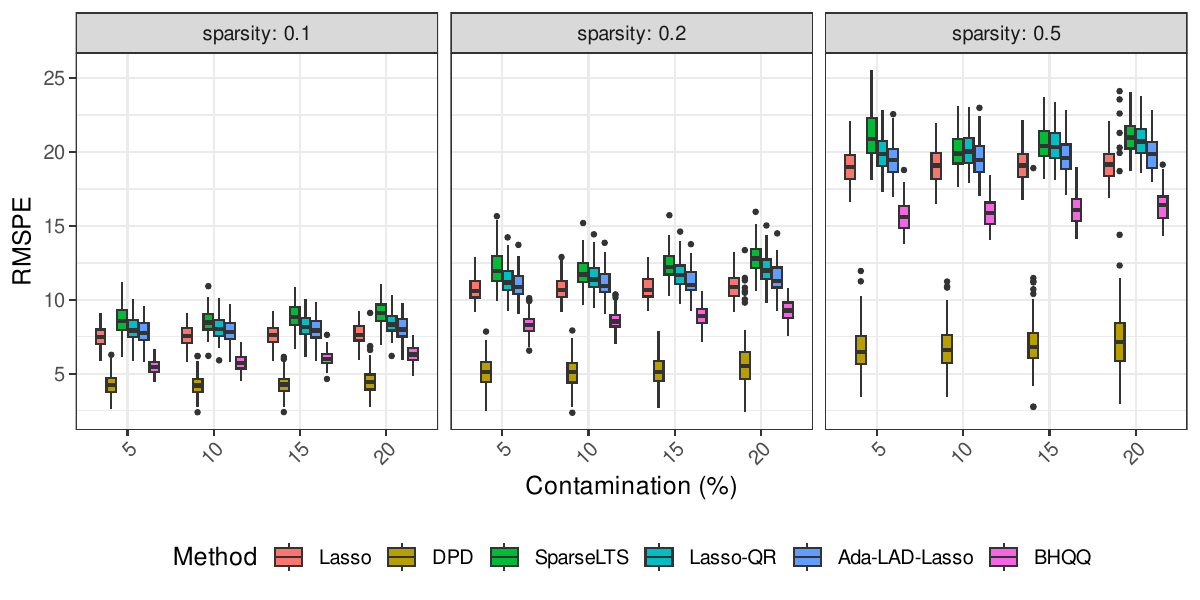}
    \caption{Boxplots of RMSPE from $B=100$ simulations with varying sparsity and contamination levels.}
    \label{fig:rmse_high}
\end{figure}

\begin{figure}[htbp]
    \centering
    \includegraphics[width=1\linewidth]{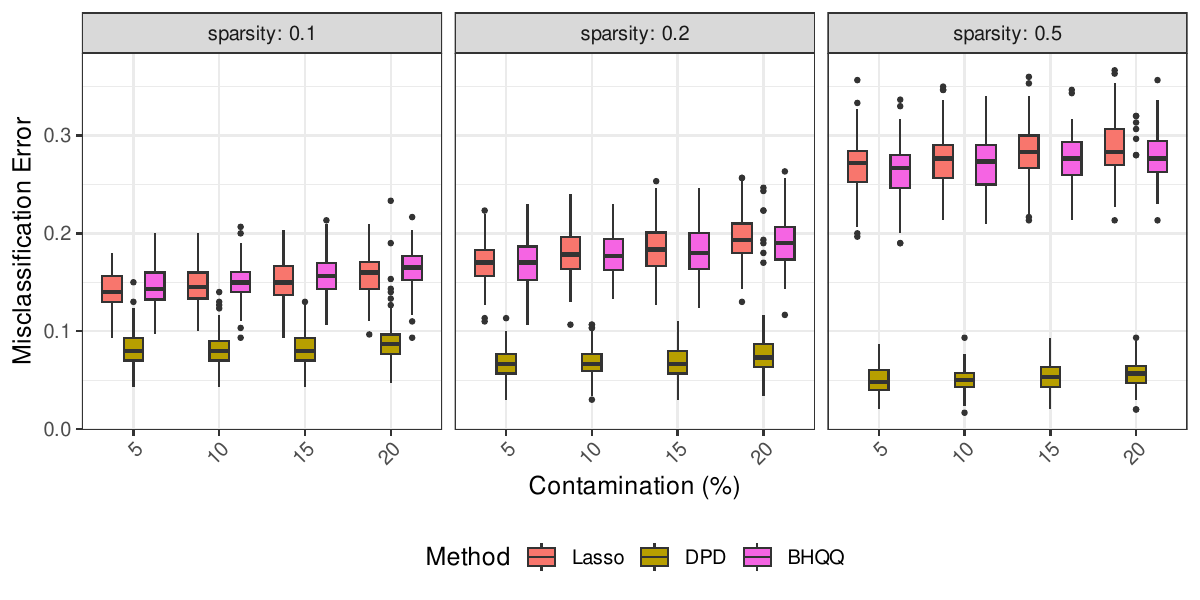}
    \caption{Boxplots of MEs from $B=100$ simulations with varying sparsity and contamination levels.}
    \label{fig:me_high}
\end{figure}

\begin{figure}[htbp]
    \centering
    \includegraphics[width=1\linewidth]{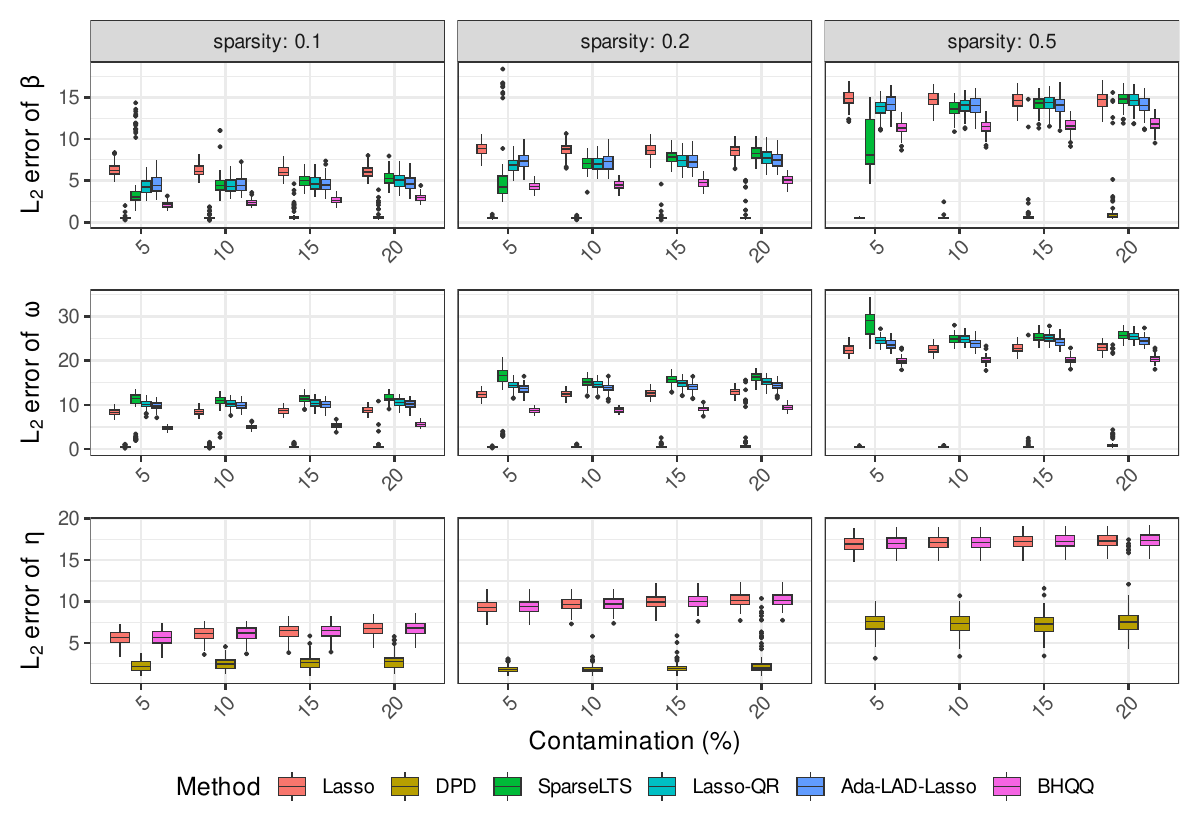}
    \caption{Boxplots of $\ell_2 E$ of $\bm \beta$, $\bm \omega$ and $\bm \eta$ from $B=100$ simulations.}
    \label{fig:betaomega_high}
\end{figure}

Figures \ref{fig:rmse_high}, \ref{fig:me_high}, and \ref{fig:betaomega_high} summarize the performance of the proposed DPD method against benchmarks in $p=50$ settings with varying sparsity and contamination.
Figure \ref{fig:rmse_high} shows that the DPD method achieves the lowest RMSPE across all sparsity levels (0.1--0.5) and contamination intensities (up to 20\%). 
While prediction errors naturally rise with sparsity for all methods, DPD maintains a clear advantage, underscoring its robustness against increased signal complexity and noise.
For classification performance (Figure \ref{fig:me_high}), DPD substantially outperforms all competitors, achieving the lowest ME. 
Its low variability also indicates greater stability than methods like Lasso and BHQQ, whose ME degrades significantly with contamination.
Figure \ref{fig:betaomega_high} evaluates parameter estimation via $\ell_2$-norm errors for $\bm \beta$, $\bm \omega$ and $\bm \eta$. 
As contamination increases, all methods are challenged, yet DPD consistently yields the most accurate estimates.
Collectively, these results demonstrate that the DPD method provides superior prediction accuracy and reliable parameter estimation in contaminated $p=50$ settings.

\spacingset{1}

\spacingset{1.8}

\section{Case Study: Lapping Process}\label{sec:real}
In this section, we apply the proposed joint models to a case study involving data from a manufacturing lapping process (see Section \ref{sec-intro}) to evaluate their empirical performance and highlight their advantages. 
The data set was previously used in \cite{Deng15} and \cite{Kang03072018}. 
It comprises two different kinds of quality responses and 450 wafer samples, each of which is represented by a set of 10 predictor variables ($x_1$–$x_{10}$). 
Total Thickness Variation (TTV), a continuous measurement $y$, and Site Total Indicator Reading (STIR), a binary measurement $z$, are included in these responses. 

Before analysis, all 10 predictor variables were standardized to have a mean of zero and a unit variance, to ensure that they are on a comparable scale. 
In the whole data set, 298 wafers have a STIR indicator of 0 (good), whereas the remaining 152 wafers have a STIR indicator of 1 (bad).  To assess model performance, the data set was partitioned into a training set of 350 samples and a test set of 100 samples.
These random partitioning processes were repeated 100 times. 
Table \ref{tab:variables} lists the ten predictor variables used in this study, categorized into controllable process settings and quality covariates measured before lapping, with detailed definitions.
The original data set of this case study can be accessed by contacting the authors. 

\spacingset{1}
\begin{table}[htbp]
\centering
\caption{Mean (standard deviation) of false positive (FP) and false negative (FN) rates for different methods in classification}
\label{tab:fp_fn}
\vspace{2mm}
\begin{tabular}{lccc}
\hline
\textbf{Metric} & \textbf{Lasso} & \textbf{DPD} & \textbf{BHQQ} \\
\hline
FP & 0.2447 (0.0638) & 0.2598 (0.0670) & 0.3935 (0.0757) \\
FN & 0.3042 (0.0539) & 0.2600 (0.0512) & 0.1250 (0.0456) \\
\hline
\end{tabular}
\end{table}
\spacingset{1.8}

\spacingset{1}
\begin{figure}[htbp]
    \centering
    \includegraphics[width=1\linewidth]{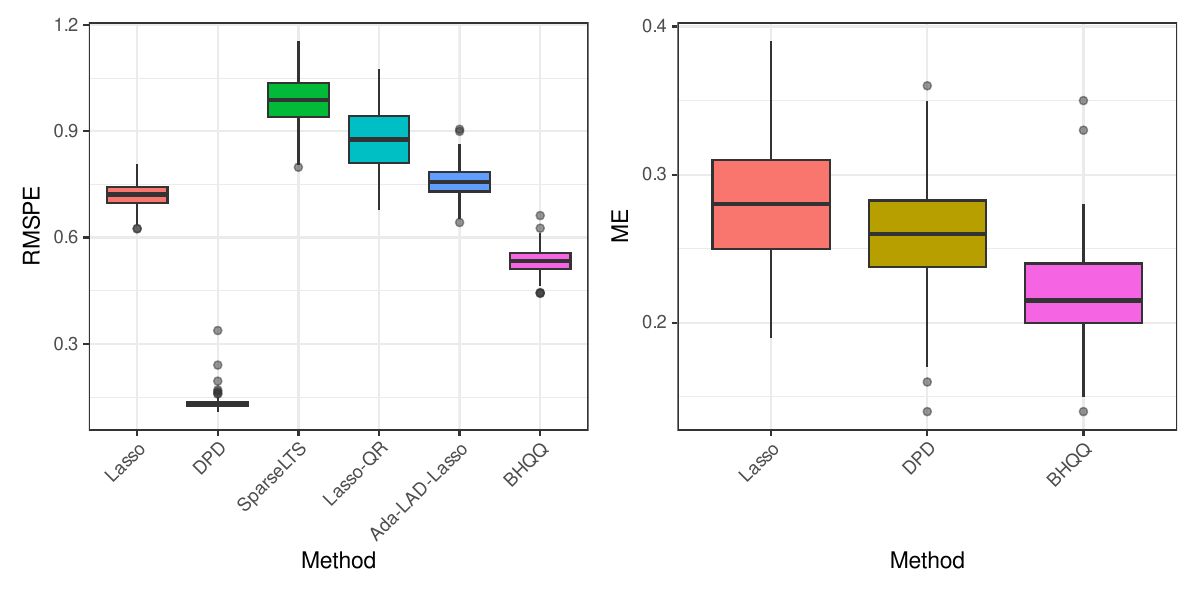}
    \caption{Boxplots of two prediction measures for lapping data for different methods}
    \label{fig:lap}
\end{figure}
\spacingset{1.8}

Figure \ref{fig:lap} presents the predictive performance for the continuous response $Y$ (RMSPE, left) and the binary response $Z$ (ME, right). 
For the continuous outcome, the proposed DPD method achieves the lowest median RMSPE (approx. 0.1) and exhibits markedly greater stability than all alternatives. 
SparseLTS and Lasso-QR perform the poorest, while Lasso and Ada-LAD-Lasso show moderate results.

For the binary classification task (ME), BHQQ attains the lowest median error, with DPD performing slightly worse. 
Lasso yields the highest median ME and the greatest variability. 
This is likely because the binary response $Z$ contains relatively few outliers, allowing BHQQ to excel on this simpler task. 
The performance of Lasso, while the worst, remains acceptable under these conditions. 
Crucially, these results highlight that our method provides a decisive advantage for the more challenging continuous prediction while remaining highly competitive for classification.

The distinct error profiles of the methods are further elucidated in Table~\ref{tab:fp_fn}, which reports the false positive (FP) and false negative (FN) rates. 
Lasso and DPD produce similar, lower FP rates than BHQQ. 
Conversely, BHQQ achieves a substantially lower FN rate. 
This trade-off stems from their underlying structures: both Lasso and DPD employ $\ell_1$-norm shrinkage, promoting sparser models. 
BHQQ, lacking such shrinkage, tends to include more predictors. 
A larger model is more sensitive to signals of the positive class, reducing the FN rate but increasing the risk of false alarms (higher FP). 
The DPD method thus offers a robust practical balance: it maintains a manageable FP rate comparable to Lasso while providing superior protection against false negatives than BHQQ.

This case study confirms the practical efficacy of the proposed DPD method for predicting wafer lapping quality. 
By delivering accurate, stable predictions and a balanced error profile across multiple data partitions, DPD demonstrates robust performance well-suited for complex industrial applications, with clear potential to enhance quality assurance in semiconductor manufacturing.

\section{Conclusion}\label{sec-conc}
This paper proposed a robust joint modeling framework that incorporates penalty terms $\ell_1$ for continuous and binary responses based on the density power divergence (DPD) criterion. 
The proposed method effectively down-weights the influence of contaminated samples and provides stable estimation even under data with outliers or mislabeled observations.
Through theoretical analysis, we established the consistency and asymptotic normality of the estimator under mild regularity conditions, ensuring a solid inferential foundation.

From a computational perspective, the proximal gradient algorithm with Barzilai–Borwein step size gives an efficient solution for the penalized DPD objective. 
The robust information criterion (RIC) was employed to select the tuning parameters in a data-driven manner, balancing model fit, robustness, and sparsity. 
Simulation results under various contamination levels demonstrate that our approach consistently achieves lower prediction errors and more accurate parameter estimation than existing methods such as Lasso, SparseLTS, and BHQQ, and the other two benchmark methods. 
In high-dimensional settings, the DPD-based estimator maintains remarkable robustness, showing both accuracy and stability as the contamination level increases.

The real wafer lapping case study further confirms the practical benefits of the proposed method. 
The proposed method delivers more reliable predictions for total thickness variation and competitive misclassification performance for the STIR indicator. 
Compared with existing alternatives, it offers a favorable trade-off between false positive and false negative rates, which is particularly important in industrial quality-control applications.

We will further investigate several directions. 
One natural extension is to accommodate more general response types, such as multiclass or ordinal outcomes. 
We also plan to develop data-driven strategies for choosing the DPD tuning parameter $\alpha$, potentially balancing robustness and efficiency in a more automatic manner instead of using a fixed $\alpha$. 

\spacingset{1.2}  
\bibliography{bibliography.bib}

\spacingset{1} 
\newpage 
\phantomsection\label{supplementary-material}
\bigskip

\setcounter{page}{1}
\setcounter{figure}{0}
\setcounter{table}{0}
\setcounter{theorem}{0}
\setcounter{algorithm}{0}
\setcounter{section}{0}

\makeatletter 
\renewcommand{\thefigure}{S\@arabic\c@figure}
\renewcommand{\thetable}{S\@arabic\c@table}
\renewcommand{\thetheorem}{S\@arabic\c@theorem}
\renewcommand{\thealgorithm}{S\@arabic\c@algorithm}
\renewcommand{\thesection}{S.\arabic{section}}
\renewcommand{\thesubsection}{S.\arabic{section}.\arabic{subsection}}
\makeatother

\begin{center}

{\Large\bf SUPPLEMENTARY MATERIAL}

\end{center}

\section{Derivation of loss function} \label{sec:Derivation_loss}
To derive $\circled{1}$ in $Q_{\alpha}$, we only need to derive $\iint f^{1+\alpha}(y,z\mid \bm x)\dd y \dd z$ for any $\bm x$ as follows. 
\begin{align*}
& \iint f^{1+\alpha}(y, z \mid \bm{x}) \dd y \dd z 
= \int \left ( f^{1+\alpha}(y, z=1 \mid \bm{x}) dy + \int f^{1+\alpha}(y, z=0 \mid \bm{x}) \right )\dd y \\
&= \int \left( \phi(y \mid \bm{x}^\top \bm{\beta}, \sigma^2) p(\bm{x} \mid \bm{\eta}) \right)^{1+\alpha} \dd y   + \int \left( \phi(y \mid \bm{x}^\top \bm{\omega}, \sigma^2) (1 - p(\bm{x} \mid \bm{\eta})) \right)^{1+\alpha} \dd y \\
&= p(\bm{x} \mid \bm{\eta})^{1+\alpha} \int \phi(y \mid \bm{x}^\top \bm{\beta}, \sigma^2)^{1+\alpha} \dd y  + (1 - p(\bm{x} \mid \bm{\eta}))^{1+\alpha} \int \phi(y \mid \bm{x}^\top \bm{\omega}, \sigma^2)^{1+\alpha} \dd y \\
&= p(\bm{x} \mid \bm{\eta})^{1+\alpha} \int \phi(\epsilon_1 \mid 0, \sigma^2)^{1+\alpha} \dd \epsilon_1 
   + (1 - p(\bm{x} \mid \bm{\eta}))^{1+\alpha} \int \phi(\epsilon_2 \mid 0, \sigma^2)^{1+\alpha} \dd \epsilon_2 \\
&= p(\bm{x} \mid \bm{\eta})^{1+\alpha} \frac{1}{(\sqrt{2\pi\sigma^2})^{\alpha}} \frac{1}{\sqrt{\alpha+1}}
   + (1 - p(\bm{x} \mid \bm{\eta}))^{1+\alpha} \frac{1}{(\sqrt{2\pi\sigma^2})^{\alpha}} \frac{1}{\sqrt{\alpha+1}} \\
&= \frac{1}{(\sqrt{2\pi\sigma^2})^{\alpha}} \frac{1}{\sqrt{\alpha+1}}
   \left[ p(\bm{x} \mid \bm{\eta})^{1+\alpha} + (1 - p(\bm{x} \mid \bm{\eta}))^{1+\alpha} \right]. 
\end{align*}

To derive the second term $\circled{2}$, we omit the scaling constant $(1+1/\alpha)(1/n)$ in the derivation. 
\[
\sum_{i=1}^n \iint g(y, z \mid \bm{x}) f^{\alpha}(y, z \mid \bm{x}_i) \dd y \, \dd z= \sum_{i=1}^n \mathbb{E}_{y,z\sim g(y,z\mid \bm x_i)} [f^{\alpha}(y, z \mid \bm{x}_i)]\\
\approx  \sum_{i=1}^n [f^{\alpha}(y_i, z_i \mid \bm{x}_i)]. 
\]
Here, we use the single point evaluation $f^{\alpha}(y_i, z_i | \bm x_i)$ to approximate the mean $\mathbb{E}_{y,z\sim g(y,z\mid \bm x_i)} [f^{\alpha}(y, z \mid \bm{x}_i)]$ because $g(\cdot |\bm x_i)$ is unknown and approximated by the empricial distribution given data $(y_i, z_i, \bm x_i)$, following the similar derivation in \cite{ghosh2013robust}. 
Therefore, 
\begin{align*}
&\sum_{i=1}^n \iint g(y, z \mid \bm{x}) f^{\alpha}(y, z \mid \bm{x}_i) \dd y \, \dd z\\
\approx & \sum_{i: z_i=1}^{n_{z=1}} \phi^{\alpha}(y_i\mid \bm x_i^\top \bm \beta, \sigma^2) p^{\alpha}(\bm x_i\mid \bm \eta) + \sum_{i: z_i=0}^{n_{z=0}} \phi^{\alpha}(y_i\mid \bm x_i^\top \bm \omega, \sigma^2)(1- p(\bm x_i\mid \bm \eta))^{\alpha}\\
=&\left(\frac{1}{\sqrt{2\pi\sigma^2}}\right)^{\alpha}\left[ \sum_{i : z_i = 1}^{n_{z=1}}
   \exp\!\left( -\frac{\alpha (y_i - \bm{x}_i^\top \bm{\beta})^2}{2\sigma^2} \right) p^{\alpha}(\bm x_i\mid \bm \eta)\right.
\\
&\left.\quad + 
   \sum_{i : z_i = 0}^{n_{z=0}}
   \exp\!\left( -\frac{\alpha (y_i - \bm{x}_i^\top \bm{\omega})^2}{2\sigma^2} \right) 
   \left( 1-p(\bm x_i\mid \bm \eta) \right)^{\alpha}\right]
\end{align*}
The resulted $Q_{\alpha}$ is $\circled{1}+\circled{2}$ and remove the scaling constant, we can obtain $Q_{\alpha}$.

\section{Proof of Theorem \ref{thm:asympt}}\label{sec:restate}
\subsection{Restatement of Theorem 2.2 in \cite{Basu1998robust}}
To ensure clarity and facilitate direct comparison with \cite{Basu1998robust}, we explicitly replace the notation used in their asymptotic theory with our notation. 
In our model, the full parameter vector is denoted by $\bm \theta =(\bm \beta^\top, \bm \omega^\top, \bm \eta^\top)$, where $\bm \beta$ and $\bm \omega$ represent the regression coefficients in different groups and $\bm \eta$ represents the classification parameter for binary output.
The joint model density is denoted by $f_{\bm \theta}(y,z \mid \bm x)$, where $y$ is the continuous response and $z$ is the binary response. 
The true data-generating density is denoted 
by $g_{\bm \theta_t}(y, z \mid \bm x)$ in our notation. 
When the model is correctly specified and $\bm \theta=\bm \theta_t$, density $g$ will be equivalent to $f_{\bm \theta}$ and density power divergence loss \eqref{eq:dpd} will be 0.

The score function is given by $\bm u_{\bm \theta}(y,z,\bm x) =\frac{\partial}{\partial \bm \theta} \log f_{\bm \theta}(y,z \mid \bm x )$. That is, 
\begin{align*}
\bm u_{\bm \beta}(y,z \mid \bm x )&=\frac{\partial}{\partial {\bm \beta}} \log f_{\bm \beta}(y,z \mid \bm x)=\frac{z}{\sigma^2}\left[y-z \bm x^\top\bm \beta-(1-z)\bm x^\top \bm \omega\right]\bm x\\
\bm u_{\bm \omega}(y,z \mid \bm x )&=\frac{\partial}{\partial {\bm \omega}} \log f_{\bm \omega}(y,z \mid \bm x)=\frac{1-z}{\sigma^2}\left[y-z \bm x^\top\bm \beta-(1-z)\bm x^\top \bm \omega\right]\bm x\\
\bm u_{\bm \eta}(y,z \mid \bm x )&=\frac{\partial}{\partial {\bm \eta}} \log f_{\bm \eta}(y,z \mid \bm x)=\left[z-p(\bm x \mid \bm \eta)\right]\bm x.
\end{align*}
According to \cite{Basu1998robust}, $\bm J$ matrix is defined as 
\begin{align*}
\bm J(\bm \theta)&=\int \bm u_{\bm \theta}(y,z|\bm x) \bm u_{\bm \theta}^\top(y, z|\bm x) f_{\bm \theta}^{1+\alpha}(y,z|\bm \theta) \dd y \dd z \\
&+\int  (\bm i_{\bm{\theta}}(y,z|\bm x)-\alpha \bm u_{\bm{\theta}}(y,z|\bm x) \bm u^\top_{\bm{\theta}}(y,z|\bm x))(g_{\bm \theta_t}(y,z|\bm x)-f_{\bm \theta}(y,z|\bm x))f^\alpha_{\bm \theta}(y,z|\bm x) \dd y \dd z, 
\end{align*}
where $\bm i_{\bm \theta}(y,z|\bm \theta) =\partial \{ \bm u_{\bm \theta}(y,z|\bm \theta)\}/ \partial {\bm \theta}^\top$. 
The paper also let $\bm K=\bm K(\bm \theta)$ be the covariance matrix of $\bm T= f_{\bm \theta}^\alpha u_{\bm \theta}$ i.e.
\[\bm K=\int \bm u_{\bm \theta} \bm u_{\bm \theta}^\top f_{\bm \theta}^{2\alpha}(y,z|\bm \theta) g_{\bm \theta_t}(y,z|\bm \theta)\dd y\dd z -\bm \xi_{\bm \theta} \bm \xi_{\bm \theta}^\top,\] 
and
\[\bm \xi_{\bm \theta}=\int \bm u_{\bm \theta}(y,z|\bm \theta) f^{\alpha}_{\bm \theta}(y,z|\bm \theta) g_{\bm \theta_t}(y,z|\bm \theta) \dd y\dd z.\]
When the true distribution $g_{\bm \theta_t}$  belongs to the parametric family $\{f_{\bm \theta}\}$ and $\bm \theta_t=\bm \theta$, the formula for $\bm J$, $\bm K$ and $\bm \xi_{\bm \theta}$ could be simplified as:
$$\bm J(\bm \theta)=\int \bm u_{\bm \theta}(y,z \mid \bm x) \bm  u_{\bm \theta}^\top(y,z \mid \bm x) f_{\bm \theta}^{1+\alpha}(y,z \mid \bm x) \dd y \,\dd z $$
$$\bm K=\int \bm u_{\bm \theta}(y,z \mid \bm x ) \bm u_{\bm \theta}^\top(y,z \mid \bm x) f_{\bm \theta}^{1+2\alpha}(y,z \mid \bm x)\dd y\, \dd z -\bm \xi_{\bm \theta} \bm \xi_{\bm \theta}^\top,$$ 
and 
$$\bm \xi_{\bm \theta}=\int \bm  u_{\bm \theta}(y,z \mid \bm x ) f^{1+\alpha}_{\bm \theta}(y,z \mid \bm x ) \dd z.$$

Next, we derive the $\bm J$ matrix via a block matrix. 
\begin{align*}
    \bm j_{\bm \beta\bm \beta^\top} & = \iint \bm u_{\bm \beta}(y,z \mid \bm x) \bm u^\top_{\bm \beta}(y,z \mid \bm x)f_{\bm \beta}^{1+\alpha}(y,z\mid \bm x ) \dd y \, \dd z \\
    & = \int \bm u_{\bm \beta}(y,z=1 \mid \bm x) \bm u^\top_{\bm \beta}(y,z=1 \mid \bm x)f_{\bm \beta}^{1+\alpha}(y,z=1\mid \bm x ) \dd y \\
    & =  \int \bm x \bm x^\top (\frac{y-\bm x^\top\bm\beta}{\sigma^2})^2 \{\phi(y\mid \bm x^\top \bm \beta,\sigma^2)p(\bm x \mid \bm \eta)\}^{1+\alpha} \dd y \\
    & = \frac{\bm x \bm x^\top}{\sigma^4}p(\bm x \mid \bm \eta)^{1+\alpha}(2\pi \sigma^2)^{-\frac{1+\alpha}{2}}\int (y-\bm x^\top \beta)^2 \exp(-\frac{(1+\alpha)(y-\bm x^\top\bm \beta)^2}{2\sigma^2})\dd y \\
    & = \frac{\bm x \bm x^\top}{\sigma^4}p(\bm x \mid \bm \eta)^{1+\alpha}(2\pi \sigma^2)^{-\frac{1+\alpha}{2}}\times \sqrt{2\pi}(\frac{\sigma^2}{1+\alpha})^{3/2} \\
    & = (2\pi\sigma^2)^{-\frac{\alpha}{2}} \sigma^{-2} (1+\alpha)^{-3/2}  p(\bm x \mid \bm \eta)^{1+\alpha} \bm x \bm x^\top
\end{align*}
Using a similar way, we could get:
\begin{align*}
\bm j_{\bm \omega\bm \omega^\top} 
& = (2\pi\sigma^2)^{-\frac{\alpha}{2}} \sigma^{-2} (1+\alpha)^{-3/2} (1-p(\bm x \mid \bm \eta))^{1+\alpha}\bm x \bm x^\top 
\end{align*}
\begin{align*}
 \bm j_{\bm \eta\bm \eta^\top} & = \iint \bm u_{\bm \eta}(y,z \mid \bm x) \bm u^\top_{\bm \eta}(y,z \mid \bm x)f_{\bm \eta}^{1+\alpha}(y,z\mid \bm x ) \dd y \, \dd z \\
 &= \int \bm u_{\bm \eta}(y,z=1 \mid \bm x) \bm u^\top_{\bm \eta}(y,z=1 \mid \bm x)f_{\bm \eta}^{1+\alpha}(y,z=1\mid \bm x ) \dd y \\
 & \quad + \int \bm u_{\bm \eta}(y,z=0 \mid \bm x) \bm u^\top_{\bm \eta}(y,z=0 \mid \bm x)f_{\bm \eta}^{1+\alpha}(y,z=0\mid \bm x ) \dd y \\
 & = (2\pi\sigma^2)^{-\alpha/2}(1+\alpha)^{-1/2}[(1-p(\bm x \mid \bm \eta))^2p(\bm x  \mid \bm \eta)^{1+\alpha}+(1-p(\bm x \mid \bm \eta))^{1+\alpha}p(\bm x  \mid \bm \eta)^{2}] \bm x \bm x^\top
\end{align*} 
\begin{align*}
    \bm j_{\bm \beta\bm \eta^\top}&= \iint \bm u_{\bm \beta}(y,z \mid \bm x) \bm u^\top_{\bm \eta}(y,z \mid \bm x)f^{1+\alpha}(y,z\mid \bm x ) \dd y \, \dd z \\
    &= \iint z \frac{(y-z\bm x ^\top \bm \beta-(1-z)\bm x ^\top \bm\omega)}{\sigma^2}\bm x (z-p(\bm x \mid \bm\eta)) \bm x^\top f^{1+\alpha}(y,z\mid x) \dd y \,  \dd z \\
    & = \frac{\bm x \bm x^\top(1-p(\bm x \mid \bm \eta))}{\sigma^2}p(\bm x \mid \bm \eta)^{1+\alpha}\int (y-\bm x ^\top \bm \beta)(\sqrt{2\pi\sigma^2})^{1+\alpha} \exp(-\frac{(1+\alpha)(y-\bm x^\top \bm \beta)^2}{2\sigma^@}) \dd y \\
    & = \bm 0
\end{align*}
Using similar tricks, we will also have the following:
\begin{align*}
     \bm j_{\bm \beta\bm \omega^\top} = \bm j_{\bm \omega\bm \eta^\top}&=\bm 0
\end{align*}
Therefore, the $\bm J$ matrix could be simplified as:
\begin{align*}
    &\bm J_{\bm \beta \bm \beta^\top}= \frac{1}{n}\sum_{i=1}^n \bm j_{\bm \beta \bm \beta^\top}=  \sum_{i=1}^n \frac{1}{n }(2\pi\sigma^2)^{-\frac{\alpha}{2}} \sigma^{-2} (1+\alpha)^{-3/2}  p(\bm x_i \mid \bm \eta)^{1+\alpha} \bm x_i \bm x_i^\top\\[0.4em]
    &\bm J_{\bm \omega \bm \omega^\top} = \sum_{i=1}^n\frac{1}{n }\bm j_{\bm \omega \omega^\top}=\sum_{i=1}^n\frac{1}{n }(2\pi\sigma^2)^{-\frac{\alpha}{2}} \sigma^{-2} (1+\alpha)^{-3/2} (1-p(\bm x_i \mid \bm \eta))^{1+\alpha}\bm x_i \bm x_i^\top \\[0.4em]
    &\bm  J_{\bm \eta \bm \eta^\top}=\frac{1}{n}\sum_{i=1}^n[(2\pi\sigma^2)^{-\alpha/2}(1+\alpha)^{-1/2}[ (1-p(\bm x_i \mid \bm \eta))^2p(\bm x_i  \mid \bm \eta)^{1+\alpha}+ (1-p(\bm x_i \mid \bm \eta))^{1+\alpha}p(\bm x_i  \mid \bm \eta)^{2}] \bm x_i \bm x_i^\top]
\end{align*}
\begin{align*}
\bm J 
&=
\begin{bmatrix}
\bm J_{\beta\beta^\top} & \bm 0 & \bm 0 \\
\bm 0 & \bm J_{\omega\omega^\top} & \bm 0 \\
\bm 0 & \bm 0 & \bm J_{\eta\eta^\top}
\end{bmatrix}
\end{align*}
To calculate the details in $\bm K$, the $\bm \xi_{\bm \theta}$ should be calculated in advance:
\begin{align*}
    \bm \xi_{\bm\beta} & =\iint \bm u_{\bm \beta}(y,z\mid \bm x)f^{1+\alpha}(y,z\mid \bm x) \dd y,\dd z \\
    & = \int \frac{y-\bm x ^\top \bm \beta}{\sigma^2}\bm x (\sqrt{2\pi\sigma^2})^{1+\alpha} \exp(-\frac{(1+\alpha)(y-\bm x ^\top \bm \beta )^2}{2\sigma^2}) \dd y \\
    & = \bm 0
\end{align*}
Using the same way, we could also get $\bm \xi_{\bm\omega} =\bm 0$.  And the details for $\bm \xi_{\bm \eta}$ is:
\begin{align*}
& \bm \xi_{\bm\eta}  = \iint\bm u_{\bm \eta}(y,z\mid \bm x)f^{1+\alpha}(y,z\mid \bm x) \dd y\,\dd z \\
& = \int (1-p(\bm x \mid \bm \eta))\bm x f^{1+\alpha}(y,z=1, \mid \bm x) p(\bm x\mid \bm \eta)^{1+\alpha} \dd y \\
& -\int p(\bm x \mid \bm \eta)\bm x f^{1+\alpha}(y,z=0, \mid \bm x) (1-p(\bm x\mid \bm \eta))^{1+\alpha} \dd y\\
& = (2\pi\sigma^2)^{-\alpha/2}(1+\alpha)^{-1/2}p(\bm x \mid \bm \eta) (1-p(\bm x \mid \bm \eta))[p(\bm x \mid \bm \eta)^\alpha-(1-p(\bm x \mid \bm \eta))^\alpha] \bm x
\end{align*}
Because $\bm \xi_{\bm \beta}=\bm \xi_{\bm \omega}=0$, the formulation of $\bm K$ could be simplified as:  
\begin{align*}
 \bm k_{\bm \beta \bm \beta} &= \iint \bm u_{\bm \beta}(y,z \mid \bm x) \bm u^\top_{\bm \beta}(y,z \mid \bm x) f^{1+2\alpha}(y,z \mid \bm x ) \dd y \, \dd z -\bm \xi_{\bm \beta}\bm \xi^\top_{\bm \beta} \\
 & = \int \bm u_{\bm \beta}(y,z=1\mid \bm x) \bm u^\top_{\bm \beta}(y,z=1\mid \bm x)  f^{1+2\alpha}(y,z=1 \mid \bm x ) \dd y   \\
 & = \int (\frac{y-\bm x^\top \bm \beta}{\sigma^2})^2 \bm x \bm x^\top \phi^{1+2\alpha}(y \mid \bm x^\top \bm \beta,\sigma^2) p(\bm x \mid \eta)^{1+2\alpha} \dd y \\
 & = \frac{p(\bm x \mid \bm \eta)^{1+2\alpha}}{(1+2\alpha)^{3/2}}\frac{1}{(2\pi)^\alpha\sigma^{2\alpha+2}} \bm x\bm x^\top 
\end{align*}
\begin{align*}
     \bm k_{\bm \omega \bm \omega} =\frac{(1-p(\bm x \mid \bm \eta))^{1+2\alpha}}{(1+2\alpha)^{3/2}}\frac{1}{(2\pi)^\alpha\sigma^{2\alpha+2}} \bm x\bm x^\top 
\end{align*}
\begin{align*}
    \bm k_{\bm \eta \bm \eta} &= \iint \bm u_{\bm \eta}(y,z\mid \bm x) \bm u_{\bm \eta}^\top(y,z\mid \bm x)f^{1+2\alpha}(y,z\mid \bm x) \dd y\,\dd z -\bm \xi_{\bm \eta}\bm \xi^\top_{\bm \eta}\\
    &=\int (1-p(\bm x \mid \bm \eta))^2\bm x\bm x^\top \phi^{1+2\alpha}(y \mid \bm x ^\top \bm \beta,\sigma^2) p(\bm x \mid \bm \eta)^{1+2\alpha} \dd y \\
    & \quad +\int (-p(\bm x \mid \bm \eta))^2\bm x\bm x^\top \phi^{1+2\alpha}(y \mid \bm x ^\top \bm \omega,\sigma^2) (1-p(\bm x \mid \bm \eta))^{1+2\alpha} \dd y  -\bm \xi_{\bm \eta}\bm \xi^\top_{\bm \eta}\\
    & = \frac{1}{\sqrt{1+2\alpha}}(2\pi\sigma^2)^{-\alpha} \{ (1-p(\bm x \mid \bm \eta))^2p(\bm x \mid \bm \eta)^{1+2\alpha}+p(\bm x \mid \bm \eta)^2(1-p(\bm x \mid \bm \eta))^{1+2\alpha}\} \bm x \bm x^\top-\bm \xi_{\bm \eta}\bm \xi^\top_{\bm \eta}
\end{align*}
And we could also get $\bm k_{\bm \beta \bm \omega}=\bm k_{\bm \beta \bm \eta}=\bm k_{\bm \omega \bm \eta}=\bm 0$.
And then, the $\bm K$ matrix could be simplified as:
\begin{align*}
    \bm K_{\bm \beta \bm \beta^\top}= \frac{1}{n} \sum_i^n \bm k_{\bm \beta \bm \beta^\top}=\frac{1}{n}\sum_i^n\frac{p(\bm x_i \mid \bm \eta)^{1+2\alpha}}{(1+2\alpha)^{3/2}}\frac{1}{(2\pi)^\alpha\sigma^{2\alpha+2}} \bm x_i\bm x_i^\top
\end{align*}
\begin{align*}
    \bm K_{\bm \omega \bm \omega^\top} =  \frac{1}{n} \sum_i^n \bm k_{\bm \omega \bm \omega^\top} =\frac{1}{n}\sum_i^n\frac{(1-p(\bm x_i \mid \bm \eta))^{1+2\alpha}}{(1+2\alpha)^{3/2}}\frac{1}{(2\pi)^\alpha\sigma^{2\alpha+2}} \bm x_i\bm x_i^\top 
\end{align*}
\begin{align*}
    & \bm K_{\bm \eta \bm \eta^\top} = \frac{1}{n}\sum_{i=1}^n[\frac{1}{\sqrt{1+2\alpha}}(2\pi\sigma^2)^{-\alpha} \{ (1-p(\bm x_i \mid \bm \eta))^2p(\bm x_i \mid \bm \eta)^{1+2\alpha}+p(\bm x_i \mid \bm \eta)^2(1-p(\bm x_i \mid \bm \eta))^{1+2\alpha}\} \bm x_i \bm x_i^\top \\
    & \quad -\bm \xi_{\bm \eta,i}\bm \xi^\top_{\bm \eta,i}], 
\end{align*} 
where
\[\bm \xi_{\bm \eta,i} =(2\pi\sigma^2)^{-\alpha/2}(1+\alpha)^{-1/2}p(\bm x_i \mid \bm \eta) (1-p(\bm x_i \mid \bm \eta))[p(\bm x_i \mid \bm \eta)^\alpha-(1-p(\bm x_i \mid \bm \eta))^\alpha] \bm x_i.\]
The $\bm K$ matrix is
\begin{align*}
\bm K
&=
\begin{bmatrix}
\bm K_{\beta\beta^\top} & \bm 0 & \bm 0 \\
\bm 0 & \bm K_{\omega\omega^\top} & \bm 0 \\
\bm 0 & \bm 0 & \bm K_{\eta\eta^\top}
\end{bmatrix}.
\end{align*}

\subsection{Assumptions (A1-A5)}\label{sec:assump}
Our model is a specific case of the density power divergence in \cite{Basu1998robust}.
Therefore, if our model also satisfies the assumptions of \cite{Basu1998robust}, it would have the asymptotic theorem. 
The assumptions for the DPD estimator are restated as:
\begin{assumption}\label{thm:assumption}
For a given $\alpha \geq 0$, the following assumptions restated from \cite{Basu1998robust} are 
\begin{itemize}
    \item[A1.] \textbf{Common Support:}  
    The distributions $f_{\bm \theta}$ and $g$ have common support $A$, so that the set on 
    which the densities are greater than zero is independent of $\bm \theta$.
    
    \item[A2.] \textbf{Differentiability:}  
    There exists an open subset $\Omega \subseteq \Theta$ containing the true parameter $\bm \theta_t$, such that for almost all $(y, z, x) \in A$ and all $\bm \theta \in \Omega$, the density $f_{\bm \theta}(y, z \mid \bm x)$ is three times differentiable with respect to $\bm \theta$. The third partial derivatives are continuous in $\bm \theta$.

    \item[A3.] \textbf{Differentiation Under the Integral:}  
    The integral $\iint f_{\bm \theta}^{1+\alpha}(y,z \mid \bm x)\, \dd y \, \dd z$ can be differentiated three 
    times with respect to $\bm \theta$, and the derivative can be taken under the integral sign.

    \item[A4.] \textbf{$\bm J$ Matrix Positive Definite:}  
    The matrix $\bm J({\bm \theta})$ is positive definite for all $\bm \theta \in \Omega$.

    \item[A5.] \textbf{Uniformly Bounded Third Derivatives:}  
    There exist functions $M_{jkl}(y,z)$ such that
    \[
    \left| \frac{\partial^3 V_{n,\theta}(y,z)}{\partial \theta_j \partial \theta_k \partial \theta_l} \right|
    \le M_{jkl}(y,z)
    \]
    for all $\bm \theta \in \Omega$, with $\mathbb{E}_g[M_{jkl}(Y,Z)] < \infty$ for all $j,k,l$, where $\mathbb{E}_g$ denotes expectation with respect to $g$.
\end{itemize}
\end{assumption}

\subsection{Verification of assumptions for our model}
 A.1–A.3 in Assumptions~\ref{thm:assumption} are readily satisfied due to the standard properties of the model densities and parameter space. Therefore, we focus on providing detailed verification for Assumptions A4 and A5, which require more technical justification. 
 
In Assumption A.4, we have to show the $\bm J$ matrix is positive definite. From Section \ref{sec:restate}, we have 
\begin{align*}
\bm J 
&=
\begin{bmatrix}
\bm J_{\beta\beta^\top} &\bm  0 & \bm 0 \\
\bm 0 & \bm J_{\omega\omega^\top} & \bm 0 \\
\bm 0 & \bm 0 & \bm J_{\eta\eta^\top}
\end{bmatrix}
\end{align*}
where \begin{align*}
    &\bm J_{\bm \beta \bm \beta^\top}=   \sum_{i=1}^n \frac{1}{n }(2\pi\sigma^2)^{-\frac{\alpha}{2}} \sigma^{-2} (1+\alpha)^{-3/2}  p(\bm x_i \mid \bm \eta)^{1+\alpha} \bm x_i \bm x_i^\top\\[0.4em]
    &\bm J_{\bm \omega \bm \omega^\top} = \sum_{i=1}^n\frac{1}{n }(2\pi\sigma^2)^{-\frac{\alpha}{2}} \sigma^{-2} (1+\alpha)^{-3/2} (1-p(\bm x_i \mid \bm \eta))^{1+\alpha}\bm x_i \bm x_i^\top \\[0.4em] 
    &\bm J_{\bm \eta \bm \eta^\top}=\frac{1}{n}\sum_{i=1}^n[(2\pi\sigma^2)^{-\alpha/2}(1+\alpha)^{-1/2}[ (1-p(\bm x_i \mid \bm \eta))^2p(\bm x_i  \mid \bm \eta)^{1+\alpha}+ (1-p(\bm x_i \mid \bm \eta))^{1+\alpha}p(\bm x_i  \mid \bm \eta)^{2}] \bm x_i \bm x_i^\top]
\end{align*}
For simplification, let us define 
\begin{align*}
     \bm J_{\bm \beta \bm \beta^\top}= \frac{1}{n}\sum_{i=1}^n  c_1(\bm x_i) \bm x_i \bm x_i^\top\\[0.4em]
    \bm J_{\bm \omega \bm \omega^\top} = \frac{1}{n}\sum_{i=1}^n  c_2(\bm x_i) \bm x_i \bm x_i^\top\\[0.4em]
    \bm J_{\bm \eta \bm \eta^\top}=\frac{1}{n}\sum_{i=1}^n  c_3(\bm x_i) \bm x_i \bm x_i^\top
\end{align*}
Because $\alpha >0 $ and $0 < p(\bm x \mid \bm \eta)<1$, so naturally we have $c_1(\bm x_i),c_2(\bm x_i),c_3(\bm x_i)>0$.
For any nonzero vector $\bm a \in \mathbb{R}^p$, 
\begin{align*}
    \bm a^\top \bm J_{\bm \beta \bm \beta^\top}\bm a= \frac{1}{n}\sum_{i=1}^n c_1(\bm x_i)(\bm a ^\top \bm x_i)^2 \geq 0
\end{align*}
And when the design matrix $\bm X$ has full rank, it will have 
\begin{align*}
    \bm a^\top \bm J_{\bm \beta \bm \beta^\top}\bm a >0, \forall a \neq 0
\end{align*}
which also implies $\bm J_{\bm \beta \bm \beta^\top}$,$ \bm J_{\bm \omega \bm \omega^\top}$ and $\bm J_{\bm \eta \bm \eta^\top}$  is positive definite. Because $\bm J$ is block diagonal with positive definite diagonal blocks, the full matrix $\bm J$ is also positive definite. And the Assumption A.4 is satisfied.

In Assumption A.5, we consider the function
\begin{align*}
V_{n,\bm \beta,\bm \omega,\bm \eta}(X_i) &= \frac{1}{\sqrt{\alpha+1}}[p(\bm x_i\mid \bm \eta)^{1+\alpha}+(1-p(\bm x_i\mid \bm \eta))^{1+\alpha}] \\
& - (1+\frac{1}{\alpha})\exp(-\frac{\alpha(y_i -\bm x_i^\top \bm \beta)^2}{2\sigma^2})p(\bm x_i \mid \bm \eta)^\alpha \bm1(z_i=1)   \\
& - (1+\frac{1}{\alpha})\exp(-\frac{\alpha(y_i-\bm x_i ^\top \bm \omega)^2}{2\sigma^2})(1-p(\bm x_i\mid \bm \eta))^\alpha  \bm1(z_i=0)
\end{align*}
and analyze its third-order derivatives with respect to the parameter components.

\begin{align*}
\frac{\partial^3 V_{n,\bm \beta,\bm \omega,\bm \eta}(X_i) }{\partial \beta_k \partial \beta_j\partial \beta_l}  & =-(1+\frac{1}{\alpha})p(\bm x_i \mid \bm \eta)^\alpha \bm 1(z_i=1) \\
& \times \exp({-\frac{\alpha(y_i-\bm x_i^\top \bm\beta)^2}{2\sigma^2})} \frac{\alpha^2}{\sigma^4}(y_i-\bm x_i^\top \bm \beta ) \left[ \frac{\alpha}{\sigma^2}(y_i-\bm x_i^\top \bm \beta)^2-3\right] x_{ij}x_{ik}x_{il}
\end{align*}
When $z_i=0$, $\frac{\partial^3 V_{n,\bm \beta,\bm \omega,\bm \eta}(X_i) }{\partial \beta_k \partial \beta_j\partial \beta_l} $ will be 0 and there will exist a function which can bound the third derivation, so we only have to show that when $z_i=1$,  there will exist a function that can also bound the third derivation.

For $z_i=1$, 
\begin{align*}
\frac{\partial^3 V_{n,\bm \beta,\bm \omega,\bm \eta}(X_i) }{\partial \beta_k \partial \beta_j\partial \beta_l}  & =-(1+\frac{1}{\alpha})p(\bm x_i \mid \bm \eta)^\alpha \exp({-\frac{\alpha(y_i-\bm x_i^\top \bm\beta)^2}{2\sigma^2})} \frac{\alpha^2}{\sigma^4}(y_i-\bm x_i^\top \bm \beta ) \left[ \frac{\alpha}{\sigma^2}(y_i-\bm x_i^\top \bm \beta)^2-3\right] x_{ij}x_{ik}x_{il}
\end{align*}
The constant $-(1+\frac{1}{\alpha})p(\bm x_i \mid \bm \eta)^\alpha\frac{\alpha^2}{\sigma^4}$ depending on the model. Since $x_{ij}, x_{ik}, x_{il}$ are assumed to be bounded, the boundedness of the third derivative reduces to showing that the function
\[
m(r) =  \exp\left(-\frac{\alpha r^2}{2\sigma^2}\right) \left[ \alpha \frac{r^3}{\sigma^2} - 3r \right], \quad \text{where} \quad r = y_i - \bm x_i^\top \bm\beta 
\] 
is dominated by an integrable function. We could focus on the case $r \geq 0 $ since $m(r)$ is an odd function and we would take the absolute value on $V$. 
Therefore, we have
\begin{align*}
    \left |m(r)\right |= \left | \exp\left(-\frac{\alpha r^2}{2\sigma^2}\right) \left[ \alpha \frac{r^3}{\sigma^2} - 3r \right]|\right | \leq \exp\left(-\frac{\alpha r^2}{2\sigma^2} \right)(\frac{\alpha}{\sigma^2} |r^3|+3|r|)
\end{align*}
And we have
\begin{align*}
    |r|^3 \exp(-ar^2) \leq (\frac{3}{2ae})^{3/2}
\end{align*}
and 
\begin{align*}
    |r| \exp(-ar^2) \leq (\frac{1}{2ae})^{1/2}, \text{where $a=\frac{\alpha}{2\sigma^2}$}
\end{align*}
So 
\begin{align*}
    |m(r)| \leq \frac{\alpha}{\sigma^2}(\frac{3}{2ae})^{3/2}+3\times (\frac{1}{2ae})^{1/2}=\frac{\alpha}{\sigma^2}(\frac{3\sigma^2}{\alpha e})^{3/2}+3(\frac{\sigma^2}{\alpha e })^{1/2}  =m_1(r)
\end{align*}
And $m_1(r)$ is a constant so $\mathbb{E}(m_1(r))<\infty$

Analogous calculations for $\boldsymbol\omega$ yield:
\begin{align*}
   \frac{\partial^3 V_{n,\bm \beta,\bm \omega,\bm \eta}(X_i) }{\partial \omega_k \partial \omega_j\partial \omega_l}  & =-(1+\frac{1}{\alpha})(1-p(\bm x_i \mid \bm \eta))^\alpha\bm 1(z_i=0) \\
   & \times \exp(-\frac{\alpha(y_i-\bm x_i^\top \bm \omega)^2}{2\sigma^2})\frac{\alpha^2}{\sigma^4} \left[ \frac{\alpha}{\sigma^2}(y_i-\bm x_i^\top \bm \omega)^3-3(y_i-\bm x_i^\top \bm \omega)\right] x_{ik} x_{ij} x_{il}
\end{align*}
Using the same trick, $\frac{\partial^3 V_{n,\bm \beta,\bm \omega,\bm \eta}(X_i) }{\partial \omega_k \partial \omega_j\partial \omega_l} $ could be bounded by a function that has a finite expectation.

For $\boldsymbol\eta$, We decompose $V_{n,\boldsymbol\beta, \boldsymbol\omega, \boldsymbol\eta}(X_i)$ into three parts with respect to $\boldsymbol\eta$:
\begin{align*}
1^\circ &= \frac{1}{\sqrt{1+\alpha}} \left[ p(\bm x_i \mid \bm\eta)^{1+\alpha} + (1-p(\bm x_i\mid \bm \eta))^{1+\alpha} \right], \\
2^\circ &= - (1+\frac{1}{\alpha})\exp\left(-\frac{\alpha (y_i - \bm x_i^\top \bm\beta)^2}{2\sigma^2}\right) p(\bm x_i\mid \bm\eta)^\alpha \bm 1(z_i=1), \\
3^\circ &= -(1+\frac{1}{\alpha}) \exp\left(-\frac{\alpha (y_i - \bm x_i^\top \bm\omega)^2}{2\sigma^2}\right) (1-p(\bm x_i\mid\bm\eta))^\alpha \bm 1 (z_i=0).
\end{align*}

The third derivative of $V_{n,\boldsymbol\beta, \boldsymbol\omega, \boldsymbol\eta}(X_i)$ with respect to $\eta_k, \eta_j, \eta_l$ is then the sum of the third derivatives of these three parts:
\[
\frac{\partial^3 V_{n,\boldsymbol\beta, \boldsymbol\omega, \boldsymbol\eta}(X_i)}{\partial \eta_k \partial \eta_j \partial \eta_l}
= \frac{\partial^3 1^\circ}{\partial \eta_k \partial \eta_j \partial \eta_l}
+ \frac{\partial^3 2^\circ}{\partial \eta_k \partial \eta_j \partial \eta_l}
+ \frac{\partial^3 3^\circ}{\partial \eta_k \partial \eta_j \partial \eta_l}.
\]
Throughout this section, we use the shorthand $p$ for $p(x_i \mid \bm\eta)$.
\begin{align*}
    \frac{\partial 1^\circ}{\partial \eta_k  }=\sqrt{\alpha+1}[p^{\alpha+1}(1-p)-p(1-p)^{\alpha+1}] x_{ik}
\end{align*}
\begin{align*}
    \frac{\partial^2 1^\circ}{\partial \eta_k \partial \eta_j } &=\sqrt{1+\alpha} \, p(1-p)  \times
     [(\alpha+1)p^\alpha(1-p)-p^{\alpha+2}-(1-p)^{\alpha+1}+(\alpha+1)p(1-p)^{\alpha+1}] x_{ij} x_{ij}
\end{align*}
\begin{align*}
    \frac{\partial^3 1^\circ}{\partial \eta_k \partial \eta_j \partial \eta_l}& =\sqrt{1+\alpha}\times p(1-p) x_{ik} x_{ij} x_{il} \times \\
    &  \{ [1-2p][(\alpha+1)p^\alpha(1-p)-p^{\alpha+2}-(1-p)^{\alpha+1}+(\alpha+1)p(1-p)^{\alpha+1}] \\
    & +\alpha(\alpha+1)p^\alpha(1-p)^2-(\alpha+1)p^{\alpha+1}(1-p)-(\alpha+2)p^{\alpha+2}(1-p) \\
    & +(\alpha+1)p(1-p)^{\alpha+1}+(\alpha+1)p(1-p)^{\alpha+2}-(\alpha+1)^2p^{\alpha+2}(1-p)  \}
\end{align*}
\begin{align*}
     \frac{\partial^3 2^\circ}{\partial \eta_k \partial \eta_j \partial \eta_l}& =-(\alpha+1)\exp(-\frac{\alpha(y-\bm x_i^\top \bm \beta)^2}{2\sigma^2})\bm 1 (z_i=1) \\
     & \quad \times p^\alpha(1-p)[(\alpha(1-p)-p)^2-\alpha p(1-p)-p(1-p)] x_{ik} x_{ij} x_{il}
\end{align*}

\begin{align*}
     \frac{\partial^3 3^\circ}{\partial \eta_k \partial \eta_j \partial \eta_l}& =-(\alpha+1)\exp(-\frac{\alpha(y-\bm x_i^\top \bm \omega)^2}{2\sigma^2})\bm 1 (z_i=0) \\
     & \quad \times p(1-p)^\alpha[(-\alpha p +1-p)(\alpha p-1-p)+\alpha(\alpha-1)p(1-p)] x_{ik} x_{ij} x_{il}
\end{align*}
\begin{align*}
    \frac{\partial^3 V_{n,\bm \beta,\bm \omega,\bm \eta}(X_i) }{\partial \beta_k \partial \beta_j\partial \eta_l}=-\frac{(\alpha+1)}{\sigma^2}\alpha p^\alpha(1-p)\bm 1 (z_i=1)\exp(-\frac{\alpha(y-\bm x_i^\top \bm \beta)^2}{2\sigma^2})[\frac{\alpha}{\sigma^2}(y_i-\bm x_i^\top \bm \beta)^2-1] x_{ik} x_{ij} x_{il}
\end{align*}
\begin{align*}
    \frac{\partial^3 V_{n,\bm \beta,\bm \omega,\bm \eta}(X_i) }{\partial \omega_k \partial \omega_j\partial \eta_l}=\frac{(\alpha+1)}{\sigma^2}\alpha p(1-p)^\alpha\bm 1 (z_i=0)\exp(-\frac{\alpha(y-\bm x_i^\top \bm \omega)^2}{2\sigma^2})[\frac{\alpha}{\sigma^2}(y_i-\bm x_i^\top \bm \omega)^2-1] x_{ik} x_{ij} x_{il}
\end{align*}
\begin{align*}
     \frac{\partial^3 V_{n,\bm \beta,\bm \omega,\bm \eta}(X_i) }{\partial \eta_k \partial \eta_j\partial \beta_l}=-\frac{\alpha(\alpha+1)}{\sigma^2}\exp(-\frac{\alpha(y_i-\bm x_i^\top \bm \beta)^2}{2\sigma^2})(y_i-\bm x_i^\top \bm \beta)\bm 1(z_i=1)p^\alpha(1-p)[\alpha(1-p)-p]x_{ik} x_{ij} x_{il}
\end{align*}
\begin{align*}
     \frac{\partial^3 V_{n,\bm \beta,\bm \omega,\bm \eta}(X_i) }{\partial \eta_k \partial \eta_j\partial \omega_l}=-\frac{\alpha(\alpha+1)}{\sigma^2}\exp(-\frac{\alpha(y_i-\bm x_i^\top \bm \omega)^2}{2\sigma^2})(y_i-\bm x_i^\top \bm \omega)\bm 1(z_i=0)p^\alpha(1-p)[\alpha(1-p)-p]x_{ik} x_{ij} x_{il}
\end{align*}
Since $0 < p(x_i\mid \bm\eta) < 1$ and $x_{ik}$ are bounded, and the exponential terms ensure rapid decay for large $|y_i|$, each term is bounded by an integrable function.
Therefore, the sum of these three third derivatives is also uniformly bounded by an integrable function, and Assumption 5 in \cite{Basu1998robust} is satisfied for the $\boldsymbol\eta$ block. 
Cross-derivatives involving different parameter blocks (e.g., $\frac{\partial^3}{\partial \beta_k \partial \omega_j \partial \eta_l}$) either vanish due to the model structure or can be bounded using the same techniques, as the relevant terms are always products of bounded covariates and exponentially decaying functions.
Therefore, for all third-order derivatives with respect to $(\boldsymbol\beta, \boldsymbol\omega, \boldsymbol\eta)$, there exist bounding functions $M_{jkl}(x)$ such that
\[
\left| \frac{\partial^3 V_{n,\boldsymbol\beta, \boldsymbol\omega, \boldsymbol\eta}(X_i)}{\partial t_j \partial t_k \partial t_l} \right| \leq M_{jkl}(x)
\]
where $t$ denotes any parameter component, and $\mathbb{E}[M_{jkl}(x)] < \infty$. This verifies that Assumption 5 in \cite{Basu1998robust} is satisfied for our model.
 
\section{The gradient of objective function}\label{sec:gradient}

Below we provide the explicit componentwise gradients for the penalized DPD objective function, as used in the proximal gradient updates in Section \ref{sec:PGA}:
\begin{align*} \label{eq:grad}
\nabla_{\boldsymbol{\beta}} Q_\alpha(\bm \beta, \bm \omega, \bm \eta)
&= -\frac{\alpha+1}{n\sigma^2} 
   \sum_{i: z_i=1}^{n_{z=1}} 
   p(\boldsymbol{x}_i \mid \boldsymbol{\eta})^{\alpha} 
   \exp\!\left[-\frac{\alpha\big(y_i - \boldsymbol{x}_i^\top \boldsymbol{\beta}\big)^2}{2\sigma^2}\right]
   \big(y_i - \boldsymbol{x}_i^\top \boldsymbol{\beta}\big)\, \boldsymbol{x}_i
\\[0.9em]
\nabla_{\boldsymbol{\omega}} Q_\alpha(\bm \beta, \bm \omega, \bm \eta)
&= -\frac{\alpha+1}{n\sigma^2} 
   \sum_{i: z_i=0}^{n_{z=0}} 
   \big(1 - p(\boldsymbol{x}_i \mid \boldsymbol{\eta})\big)^{\alpha}
     \exp\!\left[-\frac{\alpha\big(y_i - \boldsymbol{x}_i^\top \boldsymbol{\omega}\big)^2}{2\sigma^2}\right]
   \big(y_i - \boldsymbol{x}_i^\top \boldsymbol{\omega}\big)\, \boldsymbol{x}_i
\\[0.9em]
\nabla_{\boldsymbol{\eta}} Q_\alpha(\bm \beta, \bm \omega, \bm \eta)
&= \frac{\sqrt{\alpha+1}}{n} 
   \sum_{i=1}^n 
   \left[
   \frac{\boldsymbol{x}_i \exp\!\big((\boldsymbol{x}_i^\top \boldsymbol{\eta})^{\alpha+1}\big)}
        {(1+\exp(\boldsymbol{x}_i^\top \boldsymbol{\eta}))^{\alpha+2}}
   - \frac{\boldsymbol{x}_i \exp(\boldsymbol{x}_i^\top \boldsymbol{\eta})}
          {(1+\exp(\boldsymbol{x}_i^\top \boldsymbol{\eta}))^{\alpha+2}}
   \right] \notag\\
&\quad - \frac{(\alpha+1)}{n}
   \sum_{i: z_i=1}^{n_{z=1}} 
   \exp\!\left[-\frac{\alpha\big(y_i - \boldsymbol{x}_i^\top \boldsymbol{\beta}\big)^2}{2\sigma^2}\right]
   \frac{\boldsymbol{x}_i \exp(\boldsymbol{x}_i^\top \boldsymbol{\eta})^{\alpha}}
        {(1+\exp(\boldsymbol{x}_i^\top \boldsymbol{\eta}))^{\alpha+1}} \notag\\
&\quad -\frac{ (\alpha+1)}{n}
   \sum_{i: z_i=0}^{n_{z=0}} 
   \exp\!\left[-\frac{\alpha\big(y_i - \boldsymbol{x}_i^\top \boldsymbol{\omega}\big)^2}{2\sigma^2}\right]
   \frac{-\boldsymbol{x}_i \exp(\boldsymbol{x}_i^\top \boldsymbol{\eta})}
        {(1+\exp(\boldsymbol{x}_i^\top \boldsymbol{\eta}))^{\alpha+1}}  \notag\\
& = \frac{\sqrt{\alpha+1}} {n} \sum_{i=1}^n  p(\bm{x}_i \mid \bm{\eta}) \big( 1-p(\bm x_i \mid \bm \eta)\big) \left[ p(\bm x_i \mid \bm \eta)^\alpha -(1-p(\bm x_i \mid \bm \eta))^\alpha\right] \bm x_i  \notag \\
&\quad -\frac{(\alpha+1)}{n}\sum_{i: z_i=1}^{n_{z=1}}  \exp\!\left[-\frac{\alpha\big(y_i - \boldsymbol{x}_i^\top \boldsymbol{\beta}\big)^2}{2\sigma^2}\right] p(\bm x_i \mid \bm \eta)^\alpha(1-p(\bm x_i \mid \bm \eta)) \bm x_i \notag \\
&\quad +\frac{(\alpha+1)}{n}  \sum_{i: z_i=0}^{n_{z=0}}    \exp\!\left[-\frac{\alpha\big(y_i - \boldsymbol{x}_i^\top \boldsymbol{\omega}\big)^2}{2\sigma^2}\right] 
p(\bm x_i \mid \bm \eta)\big( 1-p(\bm x_i \mid \bm \eta)\big)^\alpha \bm x_i
\end{align*}

\newpage 
\section{Proximal Gradient and BB-Step Algorithms}\label{sec:alg}

\begin{algorithm}\caption{Proximal Gradient Algorithm for Minimizing  
\eqref{eq:minQ}.}\label{alg:proximal}
\noindent
{\bf Step 0} Choose the parameter $\alpha>0$ for $Q_{\alpha}$, and the regularization parameters $\lambda_1,\lambda_2,\lambda_3>0$. 
Set the parameters for the algorithm: $\delta>0$ for convergence, update factor $\zeta>1$, an integer memory size $L>0$, $(\gamma_{m,\min},\gamma_{m,\max})$ with $0<\gamma_{m,\min}<1<\gamma_{m,\max}$ for $m=1,2,3$, and a constant $\xi>0$ controlling the sufficient decrease condition in the acceptance criterion.  \\
{\bf Step 1} Initialize $\boldsymbol{\beta}^{(0)} \in \mathbb{R}^p$, $\boldsymbol{\omega}^{(0)}\in \mathbb{R}^p$, $\boldsymbol{\eta}^{(0)}\in \mathbb{R}^p$.  Initialize $\sigma^{2}$ by PSE.\\
{\bf Step 2} For $t=1,2,\dots $, do
\begin{enumerate}[leftmargin=*]
\item Choose step size $\gamma_{1,t}$, $\gamma_{2,t}$, and $\gamma_{3,t}$ according to Algorithm \ref{alg:bb}.
\item Update $\boldsymbol{\beta}$, $\boldsymbol{\omega}$ and $\boldsymbol{\eta}$:
\begin{align*}
\boldsymbol{u}_1^{(j)} &\leftarrow 
\boldsymbol{\beta}^{(t)} - \frac{1}{\gamma_{1,t}}
\nabla_{\boldsymbol{\beta}} Q(\boldsymbol{\beta}^{(t)},\boldsymbol{\omega}^{(t)},\boldsymbol{\eta}^{(t)};\sigma^{2}),
\quad \text{then} \quad
\beta_i^{(\text{temp})} = \mathrm{soft}(u_{1,i}^{(t)},\,\lambda_1/\gamma_{1,t}),\ i = 1,\ldots,p \\[0.4em]
\boldsymbol{u}_2^{(t)} &\leftarrow 
\boldsymbol{\omega}^{(t)} - \frac{1}{\gamma_{2,t}}
\nabla_{\boldsymbol{\omega}} Q(\boldsymbol{\beta}^{(t+1)},\boldsymbol{\omega}^{(t)},\boldsymbol{\eta}^{(t)};\sigma^{2}),
\quad \text{then} \quad
\omega_i^{(\text{temp})} = \mathrm{soft}(u_{2,i}^{(t)},\,\lambda_2/\gamma_{2,t}),\ i = 1,\ldots,p \\[0.4em]
\boldsymbol{u}_3^{(t)} &\leftarrow 
\boldsymbol{\eta}^{(t)} - \frac{1}{\gamma_{3,t}}
\nabla_{\boldsymbol{\eta}} Q(\boldsymbol{\beta}^{(t+1)},\boldsymbol{\omega}^{(t+1)},\boldsymbol{\eta}^{(t)};\sigma^{2}),
\quad \text{then} \quad
\eta_i^{(\text{temp})} = \mathrm{soft}(u_{3,i}^{(t)},\,\lambda_3/\gamma_{3,t}),\ i = 1,\ldots,p. 
\end{align*}
\item Check $\boldsymbol{\beta}^{(\text{temp})},\boldsymbol{\omega}^{(\text{temp})},\boldsymbol{\eta}^{(\text{temp})}$ whether satisfy the acceptance criterion: 
\begin{align*}
& h(\boldsymbol{\beta}^{(\text{temp})}, \boldsymbol{\omega}^{(\text{temp})}, \boldsymbol{\eta}^{(\text{temp})};\sigma^{2}) 
 \leq  \max \left\{ h(\boldsymbol{\beta}^{(j)}, \boldsymbol{\omega}^{(j)}, \boldsymbol{\eta}^{(j)};\sigma^{2})  - \frac{\xi }{2} \cdot \big( \gamma_{1,t}\cdot\|\boldsymbol{\beta}^{(t+1)} - \boldsymbol{\beta}^{(t)}\|_2^2 \right. \\
& \left. + \gamma_{2,t}\cdot\|\boldsymbol{\omega}^{(t+1)} - \boldsymbol{\omega}^{(t)}\|_2^2 
+ \gamma_{3,t}\cdot \|\boldsymbol{\eta}^{(t+1)} - \boldsymbol{\eta}^{(t)}\|_2^2 \big), \text{ for }j=\max(t-L, 0),\ldots, t \right\}.
\end{align*}
If satisfied, update $t \leftarrow t+1$, $\bm \beta^{(t+1)}\leftarrow \bm \beta^{(\text{temp})}$, $\bm \omega^{(t+1)}\leftarrow \bm \omega^{(\text{temp})}$, $\bm \eta^{(t+1)}\leftarrow \bm \eta^{(\text{temp})}$ 
Otherwise, update step sizes.
$\gamma_{1,t}\leftarrow \gamma_{1,t}\zeta$, 
$\gamma_{2,t}\leftarrow \gamma_{2,t}\zeta$, 
$\gamma_{3,t}\leftarrow \gamma_{3,t}\zeta$.
and go back to Step 2.2.
\item Check convergence. If 
\[
\max\left\{
\frac{\|\boldsymbol{\beta}^{(t)} - \boldsymbol{\beta}^{(t-1)}\|_2}{\|\boldsymbol{\beta}^{(t)}\|_2},\ 
\frac{\|\boldsymbol{\omega}^{(t)} - \boldsymbol{\omega}^{(t-1)}\|_2}{\|\boldsymbol{\omega}^{(t)}\|_2},\ 
\frac{\|\boldsymbol{\eta}^{(t)} - \boldsymbol{\eta}^{(t-1)}\|_2}{\|\boldsymbol{\eta}^{(t)}\|_2} 
\right\}
\le \delta,
\]
terminate the for-loop and return $\boldsymbol{\beta}^{(t)}, \boldsymbol{\omega}^{(t)}, \boldsymbol{\eta}^{(t)}$. Otherwise, continue the for-loop. 
\end{enumerate}
\end{algorithm}

\begin{algorithm}[H]
\caption{The Barzilai-Borwein Spectral Approach.}
\label{alg:bb}
\noindent
{\bf Step 0}: Receive the iteration counter $t$ and $(\gamma_{m,\min}, \gamma_{m,\max})$ from Algorithm \ref{alg:proximal}.\\
{\bf Step 1}: If $t=0$, return $\gamma_{m,t}=1$. \\
{\bf Step 2}: If $t>0$, for $m=1,2,3$\\
\begin{enumerate} 
\item Compute $\boldsymbol{\delta}^{(t)}_m$ and $\boldsymbol{g}_m^{(t)}$:
\begin{align*}
& \text{For } m=1, \boldsymbol{\delta}^{(t)}_1 = \boldsymbol{\beta}^{(t)} - \boldsymbol{\beta}^{(t-1)}, \boldsymbol{g}_1^{(t)} = \nabla Q_{\boldsymbol{\beta}}(\boldsymbol{\beta}^{(t)}) - \nabla Q(\boldsymbol{\beta}^{(t-1)}) \\
& \text{For } m=2, \boldsymbol{\delta}^{(t)}_2 = \boldsymbol{\omega}^{(t)} - \boldsymbol{\omega}^{(t-1)}, \boldsymbol{g}_2^{(t)} = \nabla Q_{\boldsymbol{\omega}}(\boldsymbol{\omega}^{(t)}) - \nabla Q(\boldsymbol{\omega}^{(t-1)}) \\
& \text{For } m=3, \boldsymbol{\delta}^{(t)}_3 = \boldsymbol{\eta}^{(t)} - \boldsymbol{\eta}^{(t-1)}, \boldsymbol{g}_3^{(t)} = \nabla Q_{\boldsymbol{\eta}}(\boldsymbol{\eta}^{(t)}) - \nabla Q(\boldsymbol{\eta}^{(t-1)})
\end{align*}
\item Compute $\gamma_{m,t}$ using either approach:
\[
\gamma_{m,t}=\frac{\langle \boldsymbol{\delta}^{(t)}_m, \boldsymbol{g}_m^{(t)} \rangle}{\langle \boldsymbol{\delta}^{(t)}_m, \boldsymbol{\delta}^{(t)}_m \rangle} 
\quad \text{or} \quad 
\gamma_{m,t}=\frac{\langle \boldsymbol{g}_m^{(t)}, \boldsymbol{g}_m^{(t)} \rangle}{\langle \boldsymbol{\delta}^{(t)}_m, \boldsymbol{g}_m^{(t)} \rangle}
\]
\item If $\gamma_{m,t}<\gamma_{m,\min}$, return $\gamma_{m,t} \leftarrow \gamma_{m,\min}$.  
If $\gamma_{m,t}>\gamma_{m,\max}$, return $\gamma_{m,t} \leftarrow \gamma_{m,\max}$.
\end{enumerate}
\end{algorithm}

\section{Setting of Tuning Parameter}\label{sec:tuning}
To ensure the reproducibility of the numerical results, we specify the tuning parameters of the proximal gradient algorithm \ref{alg:proximal} used in the simulation (Section \ref{sec-experiment} and \ref{sec:real}). 
The implementation and hyperparameter configurations for our proximal gradient are informed by the framework of \cite{yang2016sparse}. The hyperparameters used in different stages of our study are summarized in Table \ref{tab:tuning_params}:
\begin{table}[htbp]
\centering
\caption{Parameter specifications for the Algorithm \ref{alg:proximal} across different stages.}
\label{tab:tuning_params}
\begin{tabular}{@{}lllll@{}}
\toprule
Parameter & Description & Simulations & Real Case Study & Cross Validation \\ \midrule
$\alpha$ & DPD robustness parameter & 1 & 1 & 1 \\
$\delta$ & Convergence threshold  & $10^{-4}$ & $10^{-9}$ & $10^{-5}$ \\
$\zeta$ & Update factor  & 1.5 & 2 & 3 \\
$L$ & Memory size  & 8 & 5 & 5 \\
$\xi$ & Sufficient decrease control  & $10^{-4}$ & $10^{-5}$ & $10^{-5}$ \\
$\gamma_{m,max}$ & Inverse max step size bound & $10^{2}$ & $10^{30}$ & $10^{2}$ \\
$\gamma_{m,min}$ & Inverse min step size bound & $10^{-4}$ & $10^{-4}$ & $10^{-4}$ \\
$MaxIter$ & Maximum iterations & 2000 & 1000 & 200 \\
Init. & Initialization method & Lasso estimates & Lasso estimates & Lasso estimates \\ \bottomrule
\end{tabular}
\end{table}

\newpage

\section{Case Study: Predictors}

\begin{table}[htb]
\centering  
\caption{Summary of prediction variables and responses in wafer lapping study}
\label{tab:variables}
\small 
\begin{tabular}{lll}
\hline
\textbf{Variable Type}                    & \textbf{Name}                 & \textbf{Definition}                             \\ \hline
\textbf{Controllable}                     & $x_{1}$: Pressure ($N/m^{2}$) & The high pressure of the upper to lower plate   \\
\textbf{Process}                          & $x_{2}$: Rotation (Rpm)       & The rotation speed                              \\
\textbf{Variable}                         & $x_{3}$: LPTime (Sec.)        & The time for low pressure                       \\
                                          & $x_{4}$: HPTime (Sec.)        & The time for high pressure                      \\ \hline
\textbf{Quality }              & $x_{5}$: CTHK0 ($\mu m$)      & Central thickness s                     \\
\textbf{Covariates}               & $x_{5}$: CTHK0 ($\mu m$)      & Central thickness s                     \\
\textbf{before}                   & $x_{6}$: TTV0 ($\mu m$)       & Total thickness variation            \\
 \textbf{Lapping}                                         & $x_{7}$: TIRO ($\mu m$)       & Total indicator reading                \\
                                          & $x_{8}$: STIR0 ($\mu m$)      & Site total indicator reading (STIR) of wafers   \\
                                          & $x_{9}$: BOW0 ($\mu m$)       & Deviation of local warp at the center of wafers \\
                                          & $x_{10}$: WARPO ($\mu m$)     & Maximum of local warp                  \\ \hline
\textbf{Cont. Resp.} & $y$: TTV ($\mu m$)            & Continuous total thickness variation of wafers  \\
\textbf{Binary Resp.}      & $z$: STIR indicator   & Binary indicator for the conformity of STIR     \\ \hline
\end{tabular}
\end{table}

\end{document}